\documentclass[twocolumn]{aastex631}
\newcommand{\uat}[2]{#1}
\usepackage{graphicx}
\usepackage{mathtools}
\usepackage{xspace}
\usepackage{amsmath}
\usepackage[utf8]{inputenc}

\usepackage{color}
\usepackage{xurl}
\def\gsim{~\rlap{$>$}{\lower 1.0ex\hbox{$\sim$}}}
\def\lsim{~\rlap{$<$}{\lower 1.0ex\hbox{$\sim$}}}

\def\jwst{{\it JWST}\xspace}

\def\vplanet{\texttt{\footnotesize{VPLanet}}\xspace}
\def\atmesc{\texttt{\footnotesize{AtmEsc}}\xspace}

\def\eqtide{\texttt{\footnotesize{EqTide}}\xspace}

\def\stellar{\texttt{\footnotesize{STELLAR}}\xspace}

\def\vsp{\texttt{\footnotesize{VSPACE}}\xspace}
\def\mp{\texttt{\footnotesize{multi-planet}}\xspace}

\def\nbody6{\texttt{\footnotesize{NBODY6++}}\xspace}

\def\k40{{$^{40}$K}}
\def\th232{{$^{232}$Th}}
\def\co2{{CO$_2$}}

\begin{document}

\correspondingauthor{Michael Kent A. Olohoy}
    
\author[0000-0001-5489-540X,gname='Michael Kent',sname=Olohoy]{Michael Kent A. Olohoy}
    \affiliation{Department of Physics, De La Salle University, 2401 Taft Avenue, Malate, Manila, 1004 Metro Manila, Philippines}
    \affiliation{Department of Earth and Space Sciences, Rizal Technological University, Boni Avenue, Mandaluyong, 1550 Metro Manila, Philippines}
    \email{michael\_olohoy@dlsu.edu.ph}

\author[0009-0006-6305-0699,gname='Yukari Aubrey',sname=Gelani]{Yukari Aubrey D. Gelani}
    \affiliation{Department of Earth and Space Sciences, Rizal Technological University, Boni Avenue, Mandaluyong, 1550 Metro Manila, Philippines}

\author[0000-0001-6487-5445,gname=Rory,sname=Barnes]{Rory Barnes}
    \affiliation{Department of Astronomy, University of Washington, Seattle, WA 98105, USA}

\title{The Limits of Tidal Circularization and Atmospheric Escape for Mini-Neptunes and Super-Earths: The Case of TOI-2257 b}

\keywords{\uat{Star-planet interactions}{2177} --- \uat{Planetary system evolution}{2292} --- \uat{Exoplanet atmospheric evolution}{2308} --- \uat{Stellar evolution}{1599} --- \uat{Mini Neptunes}{1063} --- \uat{Super Earths}{1655}}

\begin{abstract}
    With a 35-day orbital period and $\sim$0.5 eccentricity, TOI-2257 b occupies a sparsely populated corner of transiting mini-Neptune parameter space and may lie near the orbital distance at which tidal and atmospheric escape process become negligible. To test this hypothesis, we performed simulations of the combined stellar, tidal, and atmospheric evolution and confirmed that TOI-2257 b does lie near this boundary, regardless of whether it formed with or without a primordial hydrogen envelope. In the mini-Neptune scenario, we find that all cases retain an envelope of at least 0.02 M$_\oplus$ for 8 Gyr, but that the envelope mass can be halved if it was initially larger than 0.2 M$_\oplus$. We find that tidal evolution only occurs for cases with initial envelope masses larger than 0.02 M$_\oplus$ because their radii are large enough to generate significant torques. In the super-Earth scenario, the planet could lose $\sim$10 Earth ocean masses of surface water during the $\sim$75 Myr pre-main sequence and therefore could be habitable if it formed with $\gtrsim$2x Earth's surface water mass. The predicted tidal heating surface flux tends to be similar to Io, with no cases approaching the flux required to drive a runaway greenhouse. We find that if the planet is rocky, then eccentricities larger than 0.62 today are strongly disfavored, implying that larger eccentricities could confirm the gaseous nature of the planet. These results suggest that TOI-2257 b could provide valuable constraints on mini-Neptune properties and new insight into the radius valley shape as a function of eccentricity.
\end{abstract}

\section{Introduction}

\begin{figure}[t]
    \centering
    \includegraphics[width=\linewidth]{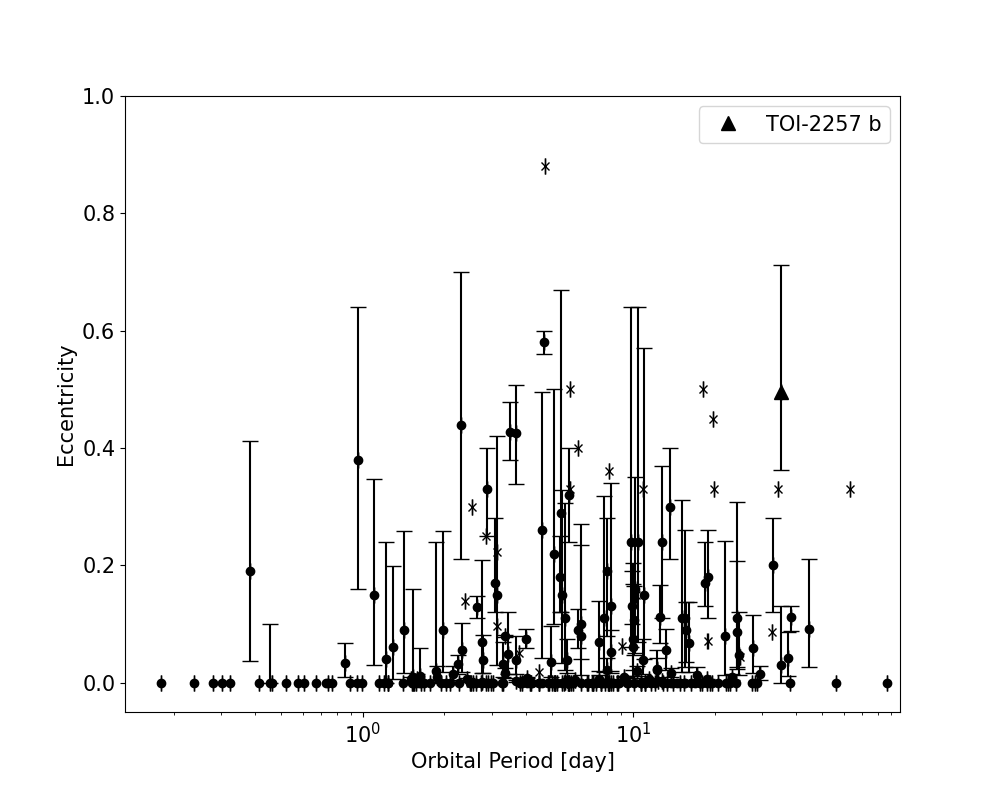}
    \caption{The eccentricities and orbital periods of planets with radii less than 4 R$_\oplus$ that transit M dwarfs, with TOI-2257 b marked as a triangle. Planets marked with crosses (×) have eccentricities reported as upper limits only. Of the 251 planets in this population, only 1 has a more eccentric orbit than TOI-2257 b (excluding planets marked with crosses), and only 8 have longer orbital periods. Data were retrieved from the NASA Exoplanet Archive \citep{Christiansen_2025} on September 4, 2026.}
    \label{fig:paramspace}
\end{figure}

TOI-2257 b is a transiting sub-Neptune-radius planet ($\sim2.2 \text{ R}_\oplus$) orbiting a 0.33 M$_\odot$ M3 star with a period $P$ of 35 days \citep{Schanche_2022}. The highly eccentric orbit of TOI-2257 b ($e = 0.5_{-0.1}^{+0.2}$) ranks it as one of the largest of the confirmed transiting exoplanets orbiting an M dwarf and its orbital period is also relatively long, see Figure~\ref{fig:paramspace}. The system's configuration also places the planet near the so-called radius valley, a relatively sparsely populated parameter space for short-period planets smaller than Neptune and larger than Earth \citep{Fulton2017}. While the boundaries of the valley are fairly well mapped in terms of planetary radius $R_\text{p}$ and $P$, the effect of eccentricity is not well constrained by the known population, especially for M dwarf host stars \citep{Wanderley_2025}. Furthermore, the orbit is at about the expected limits of where tidal effects could modify the orbit \citep{Barnes_2015} and mass loss could be significant \citep{Luger&Barnes2015,luger2015habitable}. TOI-2257 b's unique properties motivate the theoretical study presented below on the atmospheric and tidal evolution of this planet.

Radial velocity measurements of the host star failed to provide constraints on the planetary mass \citep{Schanche_2022}, so it is currently unknown if this planet is a gaseous mini-Neptune or a terrestrial super-Earth \citep{Barnes2009}. Extrapolating from planets with known masses and radii, TOI-2257 b's radius strongly favors the mini-Neptune structure \citep{Rogers2015}, but similarly sized planets are known to be terrestrial such as K2-56 b \citep{Espinoza_2016} and Kepler-145 b \citep{Xie_2014}. \cite{Schanche_2022} estimated the planetary mass using the mass-radius relation of \cite{Chen_2017}  to be $5.712_{-2.311}^{+4.288}$ M$_\oplus$, which would imply an icy/gaseous composition \citep{Zeng_2016}. The compositional ambiguity, however, remains significant; therefore we will consider both structures in our experiment. 

If the planet formed as a mini-Neptune, it may have lost significant hydrogen due to the energy from impinging XUV photons \citep{Raymond2008,Jackson2010,luger2015habitable}. The orbit-averaged instellation increases with eccentricity, so even though this planet orbits at 35 days, the large eccentricity could facilitate mass loss. Escaping hydrogen could potentially be detected as it has been from similar worlds \citep[see, e.g., ][]{Loyd2025} and its detection in this case could provide a key constraint on the efficiency of mass loss.

The orbital period also places the planet in the main sequence habitable zone (HZ) of the host star \citep{Kasting1993,Kopparapu_2013}, so it may support a habitable moon \citep{Williams1997,HellerBarnes2013,Heller2014} or, if terrestrial, be habitable itself. The low mass of the host star, however, means the star likely required approximately 200 Myr to reach the main sequence \citep{Hayashi1961,Baraffe} while planet formation likely required only a few million years \citep{Raymond2007,Lambrechts19,Clement22}. The planet may therefore have spent significant time interior to the HZ and experienced water photolysis and hydrogen escape that desiccated it \citep{Luger&Barnes2015}. However, this planet is larger and more eccentric than the planets considered in \citet{Luger&Barnes2015}, so it is not clear what conditions could result in this planet still possessing significant water today. The predicted transmission spectroscopy metric of the planet, assuming it is a mini-Neptune, is $32.7_{-14.4}^{+24.1}$, which makes it amenable to observations of the characteristics of its atmosphere via the {\it James Webb Space Telescope} (\jwst) \citep{Schanche_2022}.  Thus, a theoretical understanding of how this planet has evolved could motivate observations and/or help interpret the data. 

The planet's large eccentricity --- 0.9 would only be a 2$\sigma$ event --- is also striking in terms of tidal circularization of the orbit \citep{Rasio1996,Levrard2007,Jackson_2008,ferraz2008tidal,Barnes2017}. Tides would certainly damp the eccentricities of closer in planets \citep{Barnes_2015}, but it is unclear if that would be the case at 35-day orbital periods. Tidal circularization can also result in significant tidal heating that can hinder habitability \citep{Jackson2008habitable,Barnes2013,Driscoll&Barnes_2015} in the super-Earth assumption. Thus, a tidal evolution study could provide a better constraint on the upper limit of the eccentricity and provide further insight into planetary habitability.

In this work, we perform parameter sweeps of the tidal and atmospheric evolution of TOI-2257 b to constrain its history and current properties. We describe the physical models and initial conditions in Section~\ref{2}. The simulation results are presented in Section~\ref{3}: the XUV luminosity evolution of the host star in Section~\ref{3.1}, the atmospheric and tidal evolution of the planet assuming mini-Neptune characteristics in Section~\ref{3.2}, and the tidal evolution of the planet following the envelope-poor assumption in Section~\ref{3.3}. We then interpret the results in Section~\ref{4} and conclude in Section~\ref{5}.

\section{Methods}
    \label{2}

We used \vplanet\vplanet\footnote{Publicly available at the
\href{https://github.com/VirtualPlanetaryLaboratory/vplanet}
{VPLanet GitHub repository}.} \citep{Barnes20} to simulate the coupled planetary and stellar evolution of the system for 8 Gyr. \vplanet is a software package capable of modeling planetary systems and tracking feedback through physical processes distinguished in the software as ``modules''. The relevant stellar and planetary properties reported by \cite{Schanche_2022} are presented in Tables~\ref{Table1} and \ref{Table2}, respectively. These lists are insufficient to model the system, so we adopt parameters taken from other studies as described below. In the following subsections, we discuss the physical models used in this study and the initial conditions of the simulation.

\begin{table}
    \centering
    \caption{TOI-2257 parameters}
    \small
    \setlength{\tabcolsep}{3pt}
    \begin{tabular}{lcc}
        \hline 
        Parameter & Unit & Value \\ [0.5ex]
        \hline\hline    
        Spectral Type & & M3 \\
        B & & $16.648\pm{0.098}$ \\
        V & & $15.211\pm{0.034}$ \\
        $A_\text{v}$ & & $<0.2$ \\
        Stellar Age & Gyr & $\geq{8.0}$ \\
        Stellar Mass $M_*$ & M$_\odot$ & $0.328^{+0.021}_{-0.019}$ \\ 
        Stellar Radius $R_*$ & R$_\odot$ & $0.313\pm {0.015}$ \\
        Effective Temperature $T_\text{eff}$ & K & $3430\pm{130}$ \\ 
        Density $\rho_*$ & g/cm$^{3}$ & $15.8\pm{2.5}$ \\
        Luminosity $L_*$ & $\text{log}_{10}$(L$_\odot$) & $-1.86713^{+0.09433}_{-0.12067}$ \\
        \hline\hline
    \end{tabular}
    \label{Table1}
\end{table}

\begin{table}
    \centering
    \caption{TOI-2257 b parameters}
    \small
    \setlength{\tabcolsep}{3pt}
    \begin{tabular}{lcc}
        \hline 
        Parameter & Unit & Value \\ [0.5ex]
        \hline\hline   
        Planet radius $R_\text{P}$ & R$_\oplus$ & $2.194^{+0.113}_{-0.111}$ \\
        Planet Mass$^a$ $M_\text{P(gaseous)}$ & M$_\oplus$ & $5.712^{+4.288}_{-2.311}$ \\
        Planet Mass$^b$ $M_\text{P(solid)}$ & M$_\oplus$ & $18.311^{+3.743}_{-3.257}$ \\
        Orbital eccentricity $e$ & & $0.496^{+0.216}_{-0.133}$ \\
        Orbital Period $P$ & day & $35.189346\pm{0.000090}$ \\
        Semimajor axis $a$ & AU & $0.145\pm{0.003}$ \\
        Inclination \textit{i} & $^\circ$ & $89.786^{+0.078}_{-0.062}$ \\ 
        Equilibrium Temperature $T_{\text{eq}}$ & K & $256^{+61}_{-17}$ \\
        \hline\hline
    \end{tabular}
    \label{Table2}
    \\$^a$ Forecasted mass from \cite{Chen_2017}.
    \\$^b$ Calculated from \cite{Zeng_2016}
\end{table}

\subsection{Stellar Model}

The evolution of the host star's radius, bolometric luminosity $L_{\text{bol}}$, and effective temperature were simulated with the \stellar module \citep[see Appendix J of][]{Barnes20}. This module performs a bicubic spline interpolation of the \cite{Baraffe} model grids to compute these properties as a function of stellar mass and age. Additionally, the evolution of XUV luminosity $L_{\text{XUV}}$ follows the broken power law model of \cite{Ribas_2005}: 
\begin{equation}
    \frac{L_{\text{XUV}}}{L_{\text{bol}}} =
    \begin{cases}
        f_{\text{sat}} & t\leq t_{\text{sat}} \\
        f_{\text{sat}}\bigg( \frac{t}{t_{\text{sat}}} \bigg)^{-\beta_{\text{XUV}}}& t>t_{\text{sat}}, 
    \end{cases}
    \label{eq:broken}
\end{equation} 
where $t$ is time, $f_{\text{sat}}$ is the fraction of the bolometric luminosity in the XUV during the ``saturated'' phase $t_{\text{sat}}$, and  $\beta_{\text{XUV}}$ is the power law that describes the decay of the XUV luminosity after saturation. The value of $t_{\text{sat}}$ is difficult to constrain for M dwarfs due to poor age constraints and complications due to flaring \citep{Magaudda_2020, Johnstone_2021}. M dwarfs likely have saturation periods that exceed those of G and K dwarfs \citep{West2008,Engle2024}. Lastly, we set the simulation to integrate for 8 Gyr, the approximate age of the system, but began each simulation at 5 Myr in the \cite{Baraffe} grids to prevent errors in the interpolation method. Thus, in the following section, $t=0$ refers to 5 Myr into the \cite{Baraffe} predictions.

\subsection{Planetary Models}

Strong gravitational forces from a star deform the equilibrium shape of an orbiting planet into a prolate shape with the long axis pointing approximately toward the star \citep{Darwin_1880,GoldreichSoter1966,Penev_2024}. The equilibrium tide model assumes that, due to dissipative processes, the bulge is misaligned with respect to an imaginary line connecting the centers of mass of the star and planet, leading to a net torque that drives orbital evolution of the planet. We simulated the tidal evolution of the planet using the \eqtide module \citep[see Appendix E of][]{Barnes20}, which derives from the constant-phase-lag (CPL) model \citep{ferraz2008tidal} and constant-time-lag (CTL) model \citep{leconte2010tidal}. In this work, we utilized the CTL approach of the equilibrium tide model, where synchronous rotation was not assumed at the start of the simulations but the model allowed the planet to reach the tidally-locked state. In the CTL model, the angle between the misaligned bulge and the central line is proportional to a constant time interval $\tau$ between the passage of the perturber and the tidal bulge. We used the CTL approach as opposed to the CPL version because the latter may not be qualitatively correct for $e > 0.3$ \citep{Barnes2013}.

Based on Solar System data, giant planets are likely to dissipate less energy than rocky planets \citep{GoldreichSoter1966,AksnesFranklin2001,Henning2009}, probably by several orders of magnitude. This difference arises from a gas' natural ability to conform to the equipotential surface induced by the perturber. Despite their larger sizes, this difference in dissipation means that giant planets likely experience slower tidal evolution than rocky planets \citep{Jackson_2008,Barnes_2015}. Tidal flexing of the planetary shape also induces tidal heating, which can be significant for exoplanets on short-period orbits \citep{Jackson2008heating,Henning2009,Barnes2010,Barnes2013,Renaud2021}.

To model atmospheric loss over time, we used the \atmesc module, which accounts for some thermal and non-thermal atmospheric escape mechanisms \citep[see Appendix A of][]{Barnes20}. The heating of the upper atmosphere of the planet caused by the incident XUV flux thermally drives the hydrodynamic escape of hydrogen gas \citep{WATSON1981150,Rogers_2011}. Photoevaporative mass loss can be described by the energy-limited escape model \citep{WATSON1981150, Erkaev_2007, Lammer, Johnson_2013, Volkov_2013} and the radiation/recombination-limited escape model \citep{Murray-Clay_2009, OwenJackson, luger2015habitable, Owen_2016}, depending on the incident XUV flux. The escape rate is parameterized via the mass loss efficiency parameter, $\epsilon_{\text{XUV}}$, which encapsulates all the relevant physics related to transforming photon energy into particle velocity, such as ionization, line cooling, photochemistry, etc. However, the planetary atmosphere may also be non-thermally stripped off via Roche lobe overflow \citep{owen2016atmospheres}, which we call Bondi-limited escape \citep[e.g.,][]{Gupta,doAmaral_2022}. We assume that TOI-2257 b can experience envelope mass loss through these three processes. Other mechanisms, including core-powered mass loss \citep{Ginzburg_2016} and boil-off \citep{owen2016atmospheres,Tang_2024}, may also contribute to atmospheric escape but are not included in our simulations. Moreover, the XUV flux is treated as orbit-averaged. We adopted this secular approximation because our simulations follow the system over Gyr timescales with timesteps of 100 Myr, which is much longer than the planet’s orbital period. However, this treatment does not capture possible orbital-phase-dependent transitions between the energy-limited and radiation/recombination-limited regimes.

The planetary radius evolves following the framework established by \cite{Lopez_2012}. The model describes a planet with a rocky core and a surrounding gaseous envelope that changes over time. The convective interior follows an adiabatic temperature profile, while the upper atmosphere is assumed to become isothermal when the adiabat drops below the equilibrium temperature. The equilibrium temperature is calculated assuming a Bond albedo of 20\% and uniform redistribution of absorbed stellar radiation. The planetary radius is calculated at 10 mbar pressure level, which is consistent with the observable transit radius. The solid radius is subsequently determined using the mass-radius relationship of \cite{Sotin_2007}.

For the super-Earth case, we also consider water loss over time, which is especially relevant since the planet began interior to the HZ \citep{Luger&Barnes2015}. The \atmesc module also accounts for this process following the model of \citet{Giallucaetal_2024}, which includes energy-limited escape and diffusion-limited escape through oxygen, when applicable. In this case, the value of $\epsilon_{\text{XUV}}$ is generally lower as some energy must go into photolysis \citep{Bolmont2017}. To approximate the formation of a cold trap that sequesters water vapor below the tropopause, we only allow water loss to occur when the planet is interior to the HZ.

As desiccation progresses, it influences the accumulation of atmospheric oxygen. Planets with inefficient surface oxygen sinks can build hundreds to thousands of bars of abiotically produced $\text{O}_2$ \citep{Luger&Barnes2015, Schaeferetal_2016, Johnstone_2020, Barth2021,Carone2025}.  We assume that water loss is quenched when the HZ reaches the planetary orbit, assuming the surface temperature follows the orbit-averaged incident flux \citep{Barnes_2008}.

\cite{Turbet19} showed that the transition from a temperate climate to a post-runaway greenhouse steam atmosphere can increase the transit radius by several hundred to more than 1000 kilometers, depending on the planetary water inventory. However, we did not account for the transient radius inflation associated with a post-runaway greenhouse steam atmosphere. Since \atmesc tracks water loss and oxygen accumulation but does not explicitly model the structure of a steam atmosphere, incorporating radius inflation self-consistently is left for future work.

\subsection{Initial Conditions}

We performed a parameter sweep using 20,000 simulations each for the mini-Neptune and super-Earth scenarios with initial conditions sampled as a Gaussian of the parameters in Tables~\ref{Table1} and \ref{Table2}. We used the \vsp\footnote{Source code available at the
\href{https://github.com/VirtualPlanetaryLaboratory/vspace}
{VSPACE GitHub repository}.} and
\mp\footnote{Source code available at the
\href{https://github.com/VirtualPlanetaryLaboratory/multi-planet}
{MultiPlanet GitHub repository}.} tools \citep{Barnes20} to build and run the \vplanet simulations. We then identified those cases that matched the present-day observed parameters and analyzed them.

For the XUV evolution, we considered XUV saturation periods between 300 Myr and 3 Gyr. We adopted log$_{10}(f_{\text{sat}})=-3.14\pm0.37$ and $\beta_{\text{X}} = 1.18\pm0.31$, both varied as Gaussian, following the \cite{Jackson_2012} X-ray fits to saturated and unsaturated regimes of FGK stellar samples according to  (B - V)${_0}$. Since the host star is an M3 dwarf, the applicability of these parameters to the system should be treated with caution, as mid-M dwarfs exhibit XUV evolution histories different from those of Sun-like stars of early-M dwarfs \citep{Pineda_2021, Engle2024, Pass_2025}, while converting X-ray measurements into XUV estimates remains uncertain for M dwarfs \citep{Engle2024}. In addition, following \cite{Jackson_2012}, we assume that the X-ray luminosity drives atmospheric mass loss. However, EUV irradiation likely contributes additional high-energy flux and may therefore enhance the total atmospheric mass-loss rate \citep{King2018, King2021}. Consequently, our predicted mass loss should be regarded as a conservative estimate.

All of our simulations assume that TOI-2257 b formed in situ with no orbital migration during the protoplanetary disk phase. The duration of this phase is unknown as is the environmental conditions that the planet may have experienced. Thus, the resulting tidal and atmospheric evolution represents a possible evolutionary pathway under this formation scenario, while not accounting for potential changes in the planet’s orbital and atmospheric properties that may have occurred during its formation process.

For the mini-Neptune case, we assumed a mass of 5.712$^{+4.288}_{-2.311}$ M$_\oplus$ \citep{Schanche_2022}, considered an initial hydrogen envelope mass $M_{\text{env}}$ in the range 0.001 M$_\oplus$ -- 1 M$_\oplus$ and that 10\% -- 60\% of the incident XUV flux drives atmospheric escape. We used the distributions $0.001 \text{ s}\leq\tau_{MN}\leq0.1 \text{ s}$ and $0.01 \text{ s}\leq\tau_{\star}\leq0.1 \text{ s}$ for tidal dissipation of the planet and star, respectively, as adopted in \cite{luger2015habitable}. All these variables were sampled uniformly within the stated ranges. We then select cases from our simulations where final eccentricities and radii match the present-day observed values of TOI-2257 b.

For the super-Earth case, we calculated the planetary mass using the mass-radius relation of \cite{Zeng_2016}, assuming Earth-like composition. We assumed a mass of $18.311^{+3.743}_{-3.257}$ M$_\oplus$, an initial surface water mass $M_{\text{H}_2\text{O}}$ of 1 -- 100 terrestrial oceans (TO\footnote{1 TO = $1.39\times10^{21}$ kg of water, equivalent to the surface water content of Earth.}), and $0.05\leq\epsilon_{\text{XUV}}\leq0.3$. For $\tau$, we used the distribution  $64\text{ s}\leq\tau_{SE}\leq640\text{ s}$ \citep{Lambeck1977,Henning2009}. As with the mini-Neptune case, all these parameter distributions were sampled uniformly. The planetary radius was fixed at its present-day value throughout the simulations and radius inflation was not considered. 

\section{Results}
    \label{3}

\subsection{Stellar Evolution}
    \label{3.1}
\begin{figure}[h]
    \centering
    \includegraphics[width=\linewidth]{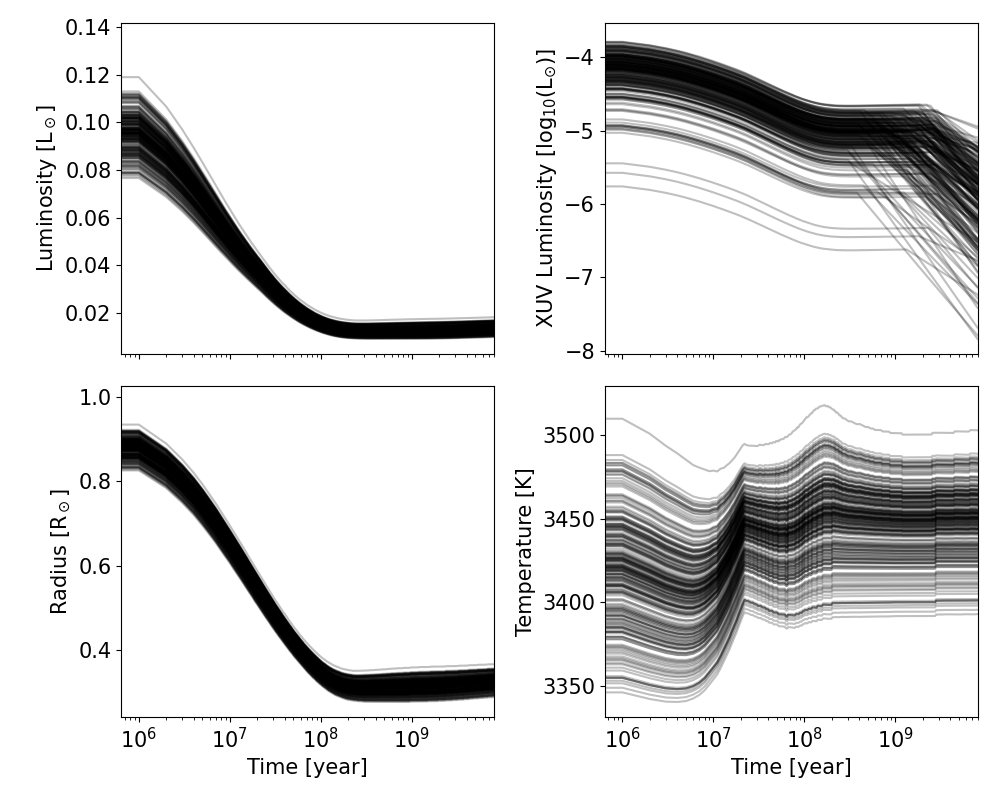}
    \caption{Two hundred randomly selected evolutions of TOI-2257's $L_{\text{bol}}$ (upper left), $L_{\text{XUV}}$ (upper right), $R_*$ (lower left), and $T_{\text{eff}}$ (lower right) after 5 Myr in the \citet{Baraffe} grid.}
    \label{fig:2}
\end{figure}

Figure~\ref{fig:2} illustrates the evolution of stellar luminosity, XUV output, radius, and effective temperature from the pre-main sequence to 8 Gyr. The star initially underwent a sharp luminosity decline and contraction in its first 100 Myr, before achieving ignition at 100--200 Myr, after which it settled into a luminosity of $\sim$0.02 L$_\odot$ and a radius  $\sim$0.3 R$_\odot$. During the pre-main sequence, the star released high-energy radiation that should remove hydrogen and/or photolyze water in the atmosphere of TOI-2257 b.  

The simulations confirm that the orbit of TOI-2257 b lies within the HZ after the star reaches the main sequence. The planet currently receives an instellation of $\sim$3400 W/m$^2$ at periapsis and $\sim$400 W/m$^2$ at apoapsis for the best fit eccentricity. Most of its revolution is spent in regions where the instellation is below the solar constant (1360 W/m$^2$), i.e., near the inner boundary of ``conservative'' habitable zone \citep{Kopparapu_2013}. Consequently, the orbit-averaged flux is only 1005 W/m$^2$, which is an influential factor for envelope retention and planetary equilibrium temperature.

\subsection{TOI-2257 b as a Mini-Neptune}
    \label{3.2}

\begin{figure}[h]
    \centering
    \includegraphics[width=\linewidth]{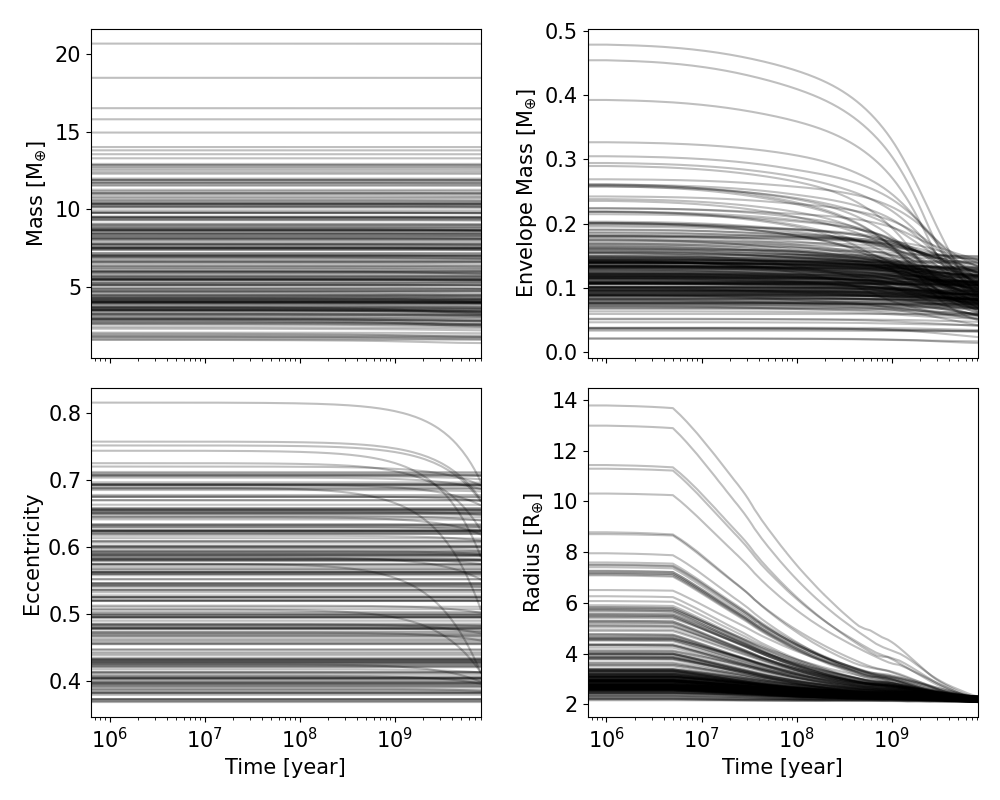}
    \caption{Evolution of $M_{\text{P}}$, $M_{\text{env}}$, $R_{\text{P}}$, and $e$ of TOI-2257 b as a mini-Neptune. The initial 20,000 simulations were filtered out to identify cases whose final eccentricities and radii were within uncertainties, as seen in Table~\ref{Table2}. There are 905 cases that were found but only $\sim$200 random cases are shown in the figure for clarity.}
    \label{fig:3}
\end{figure}

Figure~\ref{fig:3} shows example evolutions from our mini-Neptune parameter sweep. The top panels show that envelope masses below 0.2 M$_\oplus$ undergo negligible mass loss because their smaller radii and larger core mass result in more strongly bound atmospheres \citep{Lopez_2013, Lopez_2014}, but a significant decrease is observed as the planet begins with a higher initial $M_{\text{env}}$. Nevertheless, we observed that the planet retained at least 0.02 M$_\oplus$ of envelope mass for all simulations. The planetary radius $R_{\text{P}}$ decreased significantly over 8 Gyr primarily due to gravitational contraction \citep{Lopez_2012}. The radius was not forced to converge to the observed present-day value; rather, it evolved according to the adopted physical model. After the simulations were completed, we retained only those cases whose final radii were consistent with the observed radius.

In most cases, the eccentricity remains constant unless the initial eccentricity is larger than 0.7. The corresponding inward migration is also negligible, producing little change in the irradiation received by the planet and thus in its atmospheric-escape rate. However, some examples still begin circularizing due to a large envelope at lower eccentricity, which influences their radii. Planets with larger initial radii experience more rapid eccentricity damping, since $\dot{e}\propto R^5_\text{P}$ \citep{Darwin_1880, GoldreichSoter1966}. As atmospheric escape reduces $R_{\text{P}}$ over time in these cases, the rate of eccentricity evolution decreases correspondingly \citep{Jackson2010}.

\begin{figure}[h]
    \centering
    \includegraphics[width=\linewidth]{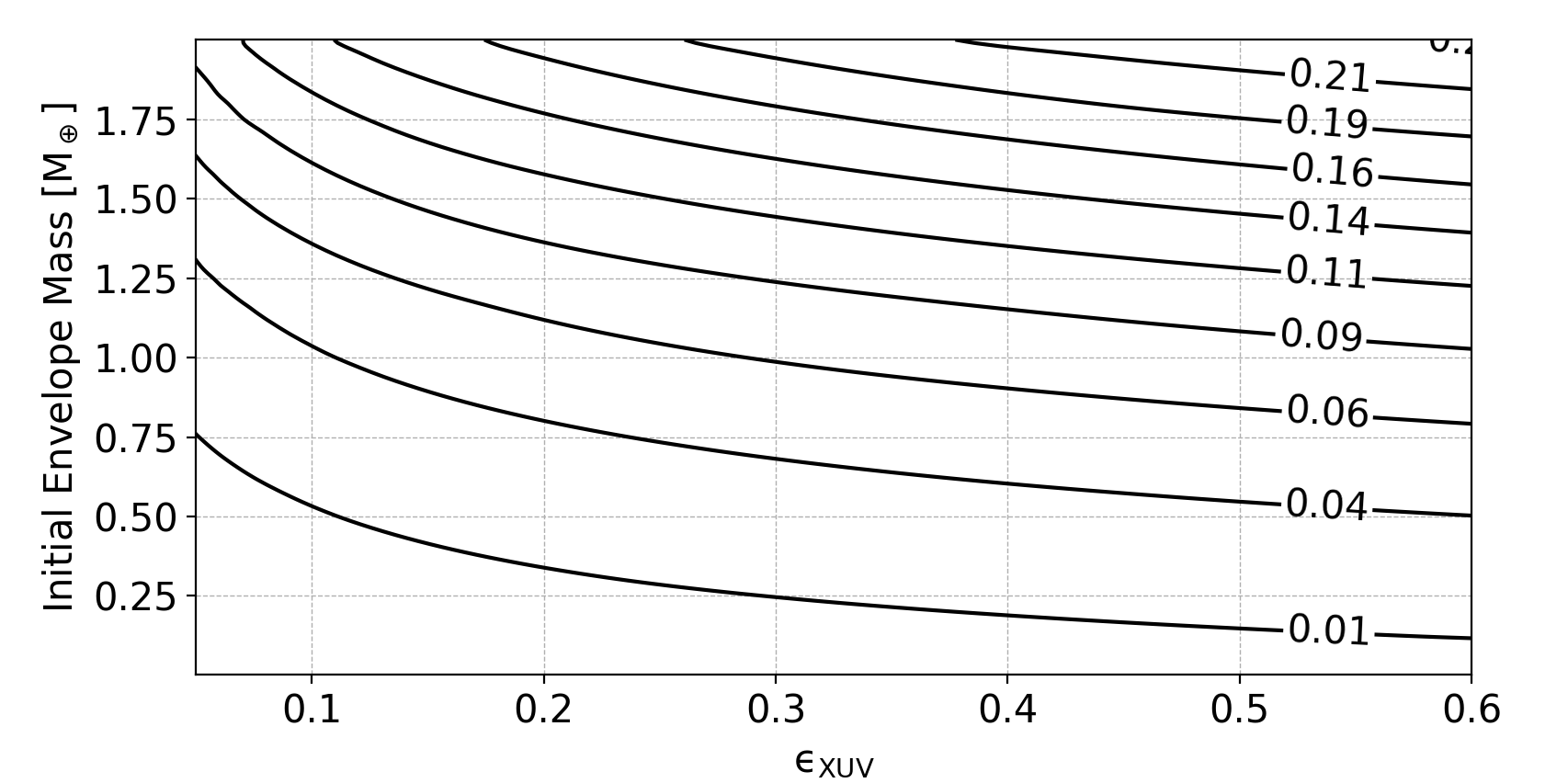}
    \caption{Change in envelope mass $\Delta M_{\text{env}}$ $\left(\text{M}_\oplus\right)$ of TOI-2257 b (mini-Neptune case) as a function of initial $M_{\text{env}}$ and $\epsilon_{\text{XUV}}$ of range [0.05, 1]. Contours represent $\Delta M_{\text{env}}$ after evolution under photoevaporative mass loss. Higher $\epsilon_{\text{XUV}}$ leads to greater atmospheric erosion, especially when the planet has a low initial $M_{\text{env}}$.}
    \label{fig:4}
\end{figure}

To further explore mass loss, we performed a parameter sweep over $\epsilon_{\text{XUV}}$ and initial envelope mass ($M_{\text{env}}$) with eccentricity held fixed at the best fit value of $e = 0.496$. To explore the behavior in the high-envelope-mass regime, we extended the initial $M_{\text{env}}$ up to 2 M$_\oplus$. We also relax our upper limit on $\epsilon_{\text{XUV}}$ to see what the maximum mass loss rates could be. Figure~\ref{fig:4} shows that even in this parameter space, the resulting mass loss is minimal. Moreover, we find that atmospheric escape is negligible for initial envelope masses $< 1 \text{ M}_\oplus$. The planet retains its atmosphere across a range of lower initial envelope masses, even under conditions of high absorption efficiency.

\subsection{TOI-2257 b as a Super-Earth}
    \label{3.3}

\begin{figure}[h]
    \centering
    \includegraphics[width=\linewidth]{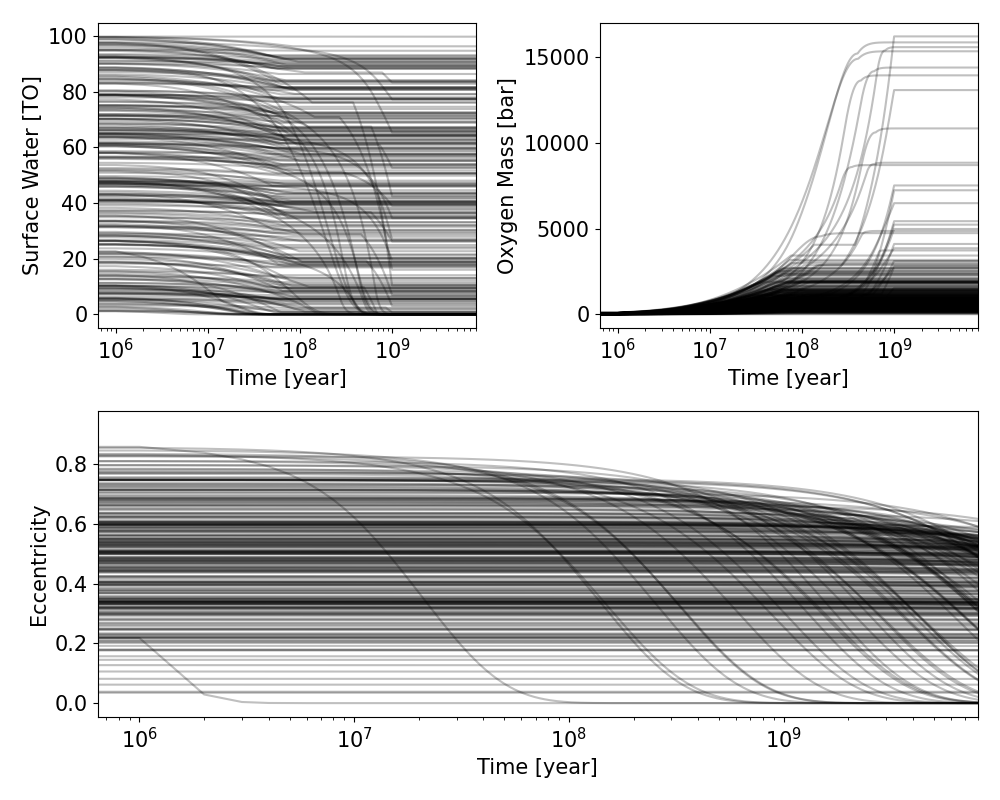}
    \caption{Tidal evolution and volatile inventory of TOI-2257 b as a super-Earth from 0.6 Myr to 8 Gyr for 200 randomly selected cases under the super-Earth assumption. \textit{Upper Left}: Surface water mass. The cases that show multiple stages of water loss crossed the inner edge of the habitable zone twice due to the combination of stellar evolution and tidal decay of the orbit \citep{Barnes_2008}. \textit{Upper Right}: Oxygen mass in the atmosphere. \textit{Bottom}: Eccentricity. Some cases circularized even with small initial eccentricity, which can be attributed to their relatively small semimajor axis.}
    \label{fig:5}
\end{figure}

Figure~\ref{fig:5} shows example evolutions for $e$, $M_{\text{H}_2\text{O}}$, and oxygen mass under the super-Earth assumption. The results show two distinct behaviors for eccentricity evolution. Many cases retain moderate eccentricities between 0.2 and 0.6, but can decline afterwards. However, not all cases show significant change, especially for lower eccentricities because tidal dissipation scales with eccentricity squared. Cases with higher initial eccentricities, particularly $e = 0.8$ tend to exhibit stronger and more rapid eccentricity damping, with some eventually circularizing.

\begin{figure}[h]
    \centering
    \includegraphics[width=\linewidth]{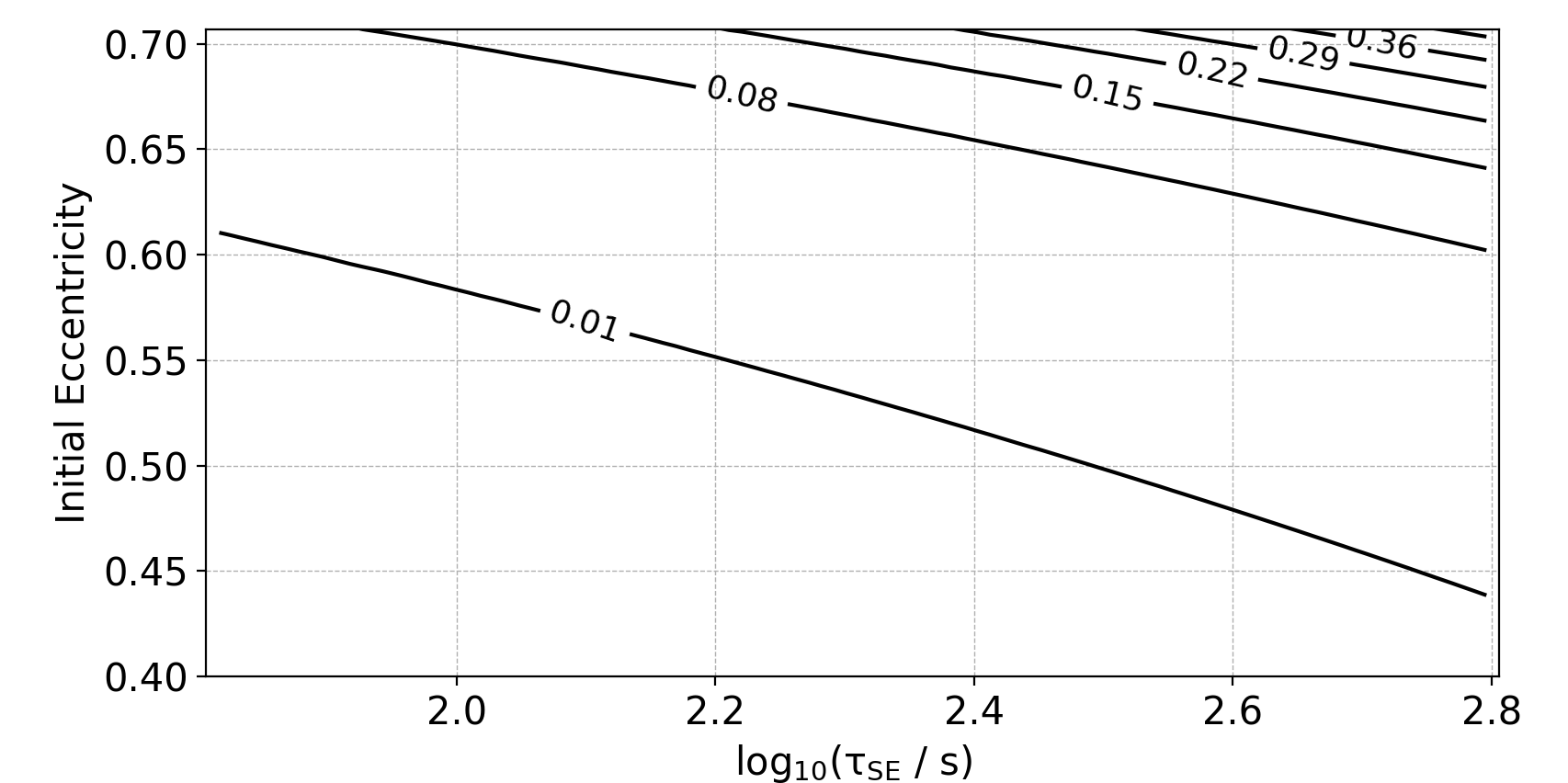}
    \caption{Change in eccentricity $\Delta e$ of TOI-2257 b (super-Earth case) as a function of initial $e$ and log-transformed tidal $\tau$. The contour lines represent $\Delta e$ as the planet experiences tidal effects throughout the simulation. Higher $\tau$ values result in greater eccentricity decay.}
    \label{fig:6}
\end{figure}

Figure~\ref{fig:6} presents $\Delta e$ as a function of $\tau$ and initial eccentricity, with planetary mass held constant at $18.311 \text{ M}_\oplus$ following the \cite{Zeng_2016} mass-radius relation. As expected, this result shows that eccentricity damping increases with higher initial eccentricity and higher $\tau$, with the largest $\Delta e$ occurring in systems that begin with moderate to high initial eccentricity and experience strong tidal dissipation. For this system, initial eccentricities lower than 0.6 appear to result in a decrease of only 0.01, suggesting the planet's eccentricity is near the limit to where tidal damping becomes negligible.

We also examined the predicted tidal heating in the planet, focusing on the surface energy flux. Approximately 58\% of the data fall below 2 W/m$^2$ of surface energy flux due to tidal heating, which is a regime comparable to the tidal heating experienced by Io \citep{Veeder1994}, while 77\% fall below 10 W/m$^2$. The largest value in our simulations is 116.8 W/m$^2$. Thus, all cases fell comfortably below the runaway greenhouse surface flux of $\sim$300 W/m$^2$ \citep{Abe1993,Kasting1993,Goldblatt2013}, indicating that the planet is not in danger of entering a tidally driven runaway greenhouse \citep{Barnes2013,Driscoll&Barnes_2015}. 

The results in the upper left panel of Figure~\ref{fig:5} show that many cases retain most of their initial surface water over time. However, some cases experience near-complete desiccation. Significant water loss occurs between $10^8$ and $10^9$ years, and is not characterized by a uniform rate. Some cases undergo periods of rapid loss followed by periods of much slower evolution. The results in the top right panel of Figure~\ref{fig:5} show a steady increase in oxygen levels, some exceeding 15000 bar, with significant oxygen accumulation occurring broadly during the same period as substantial water loss. These results are consistent with previous results \citep{Luger&Barnes2015}, and suggest that atmospheric spectra of the planet could reveal a large oxygen-rich atmosphere.

\begin{figure}[h]
    \centering
    \includegraphics[width=\linewidth]{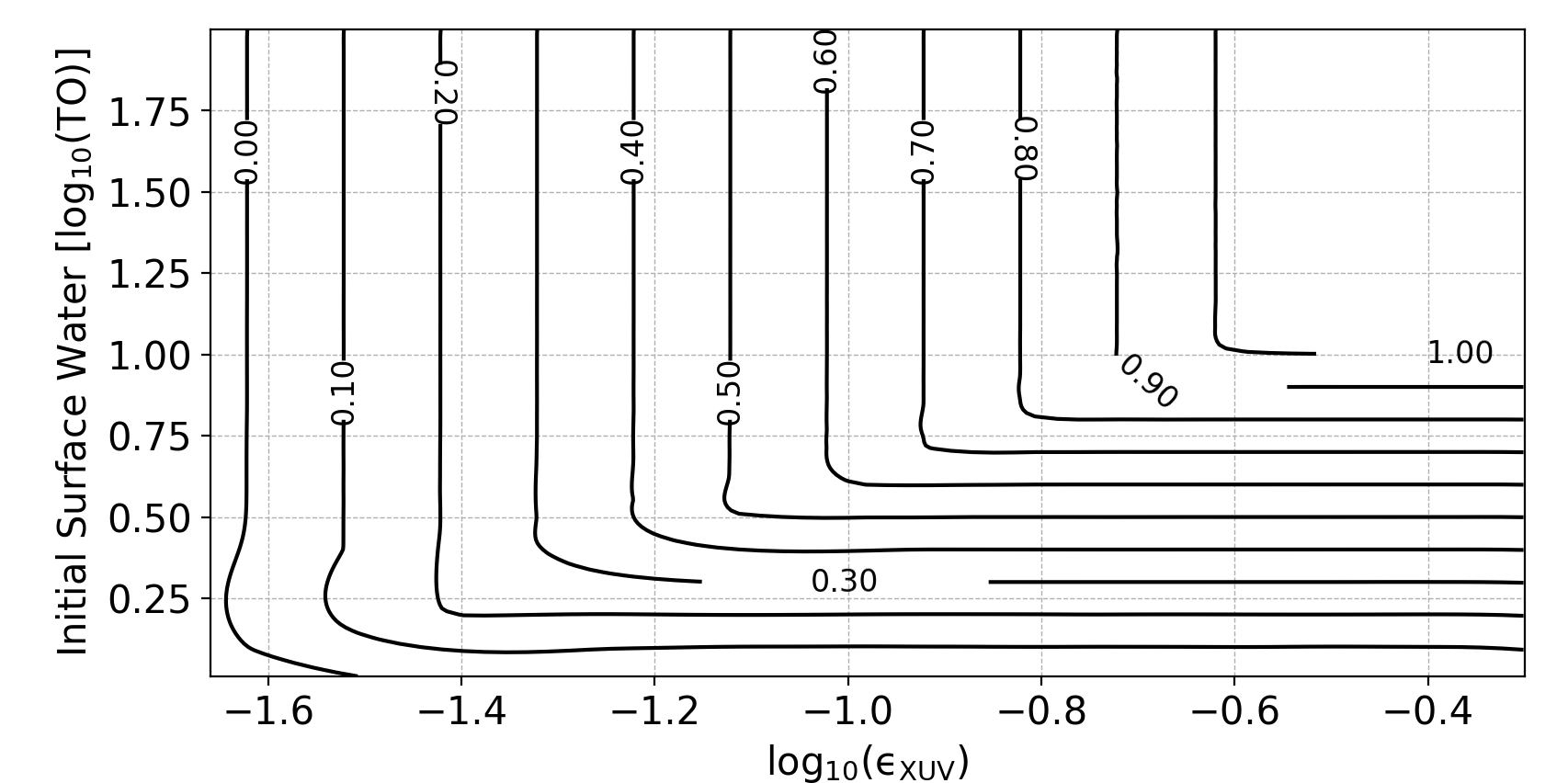}
    \caption{Change in surface water $\Delta M_{\text{H}_2\text{O}}$ in TO of TOI-2257 b (super-Earth case) as a function of initial $M_{\text{H}_2\text{O}}$ and $\epsilon_{\text{XUV}}$. All the variables are logarithmic, including the $\Delta M_{\text{H}_2\text{O}}$ contour lines. The vertical segments correspond to scenarios in which photoevaporation of surface water has not exceeded the initial water inventory and surface water remains present. The horizontal segments indicate desiccation of surface water.}
    \label{fig:7}
\end{figure}

Figure~\ref{fig:7} shows a contour map of water loss from TOI-2257 b as a function of the initial surface water abundance and $\epsilon_{\text{XUV}}$, while other parameters are held constant (i.e., $e = 0.496$, $\mathrm{log_{10}}\left(f_{\text{sat}}\right) = -3.14$, $\beta_{\text{XUV}} = 1.18$, and an XUV saturation period of $10^{8.21}$ years). As expected, higher XUV absorption efficiencies lead to complete surface water loss, as indicated by the horizontal curves. We find that no surface water is lost for absorption efficiencies $<0.007$, as the efficiency of converting incident XUV energy into atmospheric escape is too low to remove a significant amount of water before the planet enters the habitable zone, where our model assumes water loss is quenched by the formation of a cold trap.

\begin{figure}[h]
    \centering
    \includegraphics[width=\linewidth]{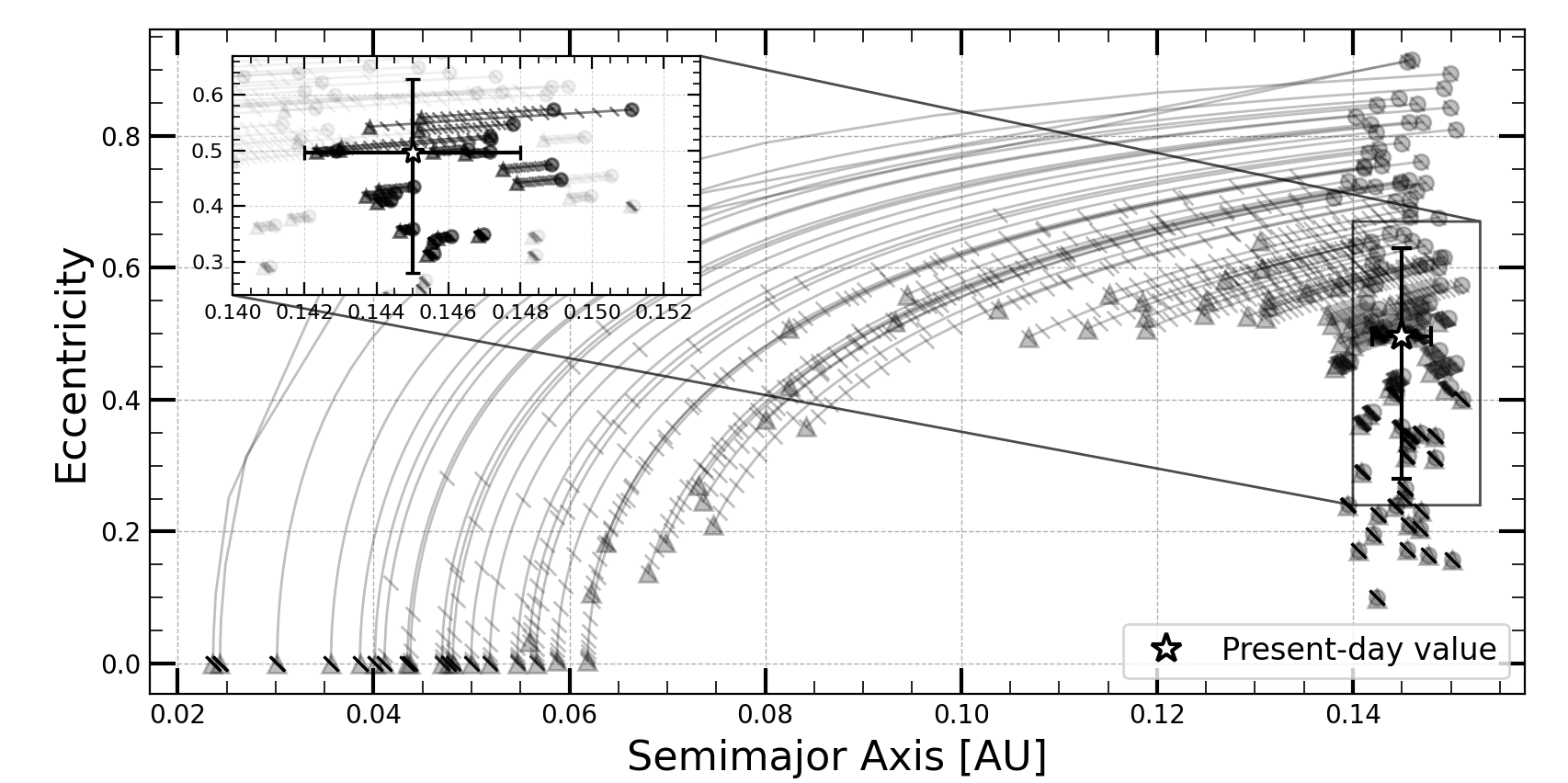}
    \caption{Evolution of the semimajor axis and eccentricity for the super-Earth case, showing only 100 cases to avoid cluttering the figure. The filled circle denotes the initial semimajor axis and eccentricity, while the triangle denotes their final values. Tick marks along the trajectory indicate 500 Myr intervals during the tidal evolution. The white star represents the present-day \textit{e} and \textit{a} of TOI-2257 b with their associated uncertainties. Tracks terminating within the present-day \textit{e} and \textit{a} uncertainty ranges are darkened. Note that there are no tick marks obscured by the upper left smaller panel.}
    \label{fig:8}
\end{figure}

Figure~\ref{fig:8} illustrates the coupled evolution of the semimajor axis and eccentricity. Significant orbital evolution can be observed for cases with higher initial eccentricities ($e\geq0.8$), which undergo both substantial inward migration and circularization. These cases migrate from initial semimajor axes of $\sim0.14$ AU to final values around $0.04$ AU while their eccentricities decrease considerably. Much of this evolution occurs within the first 500 Myr, reflecting the stronger tidal dissipation associated with highly eccentric orbits. For initial eccentricities around 0.7, significant inward migration is still present, although eccentricity damping continues beyond 500 Myr, with some cases retaining appreciable eccentricity. For intermediate initial eccentricities ($0.4\leq e\leq0.6$), both inward migration and eccentricity damping become progressively weaker. Below $e=0.4$, the orbits exhibit little to no significant inward migration or eccentricity damping over the simulated evolution.

For the most extreme cases, the simulations show that planets with high initial eccentricities can undergo substantial inward migration, evolving from initial semimajor axes of approximately $0.14$--$0.15$ AU to final values of $\sim0.04$--$0.06$ AU, corresponding to inward migration of roughly $0.08$--$0.11$ AU. This substantial migration occurs alongside strong eccentricity damping and explains why planets initialized at lower eccentricities exhibit relatively little orbital evolution, whereas initially highly eccentric planets can continue to evolve even after reaching similarly low eccentricities. Tidal evolution depends on both eccentricity and semimajor axis; thus, by the time an initially highly eccentric planet reaches a lower eccentricity, it has already migrated closer to the host star, where tidal effects are stronger. Consequently, a planet that evolves to a given eccentricity at a smaller semimajor axis can experience stronger continued tidal evolution than a planet initialized at the same eccentricity but at a larger orbital distance.

The evolutions shown in Figure~\ref{fig:8} split into 3 categories based on initial eccentricity: 1) very high eccentricity cases ($e \gtrsim 0.65$) fully circularize within the age of the system, 2) high eccentricity cases ($0.4\lesssim e\lesssim0.65$) experience some orbital evolution but do not fully circularize, and 3) low-to-moderate eccentricity cases ($e \lesssim 0.4$) that experience negligible tidal evolution. The first case arises due tidal evolution's strong, but different, dependencies on eccentricity and semi-major axes in the CTL model: the eccentricity evolution rate scales as $e^2 a^{-6.5}$, whereas the semimajor-axis evolution rate scales as $e^2 a^{-5.5}$ \citep{leconte2010tidal}. At very high eccentricity, the $e^2$ dominates, which drives rapid circularization and orbital decay. However, as the eccentricity term weakens, the semi-major axis term increases and the orbit shrinks and completely circularizes. For case 3, both the eccentricity and semi-major axis terms are small enough that almost no orbital evolution occurs. The narrow case 2 is the transitional region where orbital circularization occurs on a longer time scale and/or can stall out. This behavior should be expected at the boundary between tidally evolving and non-tidally evolving worlds because both $a$ and $e$ terms should be on the cusp of driving significant orbital evolution on Gyr timescales.

\section{Discussion}
    \label{4}

\subsection{The Mini-Neptune Case}

We find that when TOI-2257 b is assumed to be a mini-Neptune, which is the most likely case, then it experienced minimal mass loss and orbital circularization. In conjunction with past work \citep[e.g., ][]{Jackson2010,Lopez_2013,Lopez_2014}, these results suggest that this planet lies at the edge of where star-planet interactions are significant for gaseous worlds. This hypothesis was confirmed in Figures~\ref{fig:4} and \ref{fig:6}, which map out where similar planets would experience negligible to significant evolution to help guide future interpretations of planet discoveries.

TOI-2257 b retains a substantial hydrogen-rich envelope throughout its 8 Gyr history despite extended exposure to XUV luminosity during its early evolution. The planet can experience significant atmospheric escape driven by photoevaporative loss, but avoids complete atmospheric stripping because planets with small $M_\text{env}$ have relatively higher surface gravities at the XUV absorption radius, which halts mass loss \citep{Lopez_2014,luger2015habitable,Owen_2016}. As shown in Figure~\ref{fig:3}, no successful simulation ends with an envelope mass less than 0.02 M$_\oplus$. In other words, thermal escape cannot transform TOI-2257 b from a mini-Neptune into a super-Earth. The results support the interpretation that TOI-2257 b may currently host a detectable atmosphere \citep{Schanche_2022}, making it a promising target for atmospheric characterization with \jwst.

Figure~\ref{fig:3} reveals that the most likely envelope mass for this planet is 0.1 M$_\oplus$, or about 1\% of the total mass. This apparent equilibrium value appears to be due to the physical and orbital parameters of this particular system -- we did perform simulations of different configurations and found no preferred envelope mass. While a comprehensive study of this specific phenomenon was beyond the scope of this experiment, our results suggest that mini-Neptunes with large initial envelopes and that experience modest mass loss may converge to a specific envelope mass.

If gaseous, TOI-2257 b undergoes some tidal circularization for the largest initial total masses sampled in our simulations, whereas lower-mass cases retain substantial eccentricity over the integration time (see Figure~\ref{fig:3}). Some simulations also show circularization from lower initial eccentricities. These results suggest that the planet could currently be undergoing atmospheric mass loss and orbital circularization, although we emphasize that only a small fraction of our simulated parameter space produced this outcome.

TOI-2257 b could represent a valuable planet in terms of mapping out the boundaries of the so-called radius valley \citep{Fulton2017}. Most previous work has not included a planet with such a large eccentricity, suggesting this planet could open up a new dimension to constrain mass loss mechanisms, e.g. photoevaporation vs. core-powered mass loss \citep{OwenWu2013,LehmerCatling2017,Ginzburg2018,Rogers2021}. Thus, we encourage observations to confirm the presence of a hydrogen envelope, which could be through either the direct detection of hydrogen or by proving the eccentricity is larger than 0.62 (see Figures~\ref{fig:3} and \ref{fig:5}). 

\subsection{The Super-Earth Case}

Our simulations show that if TOI-2257 b formed as a rocky planet, its eccentricity and atmosphere could have undergone substantial evolution. The results described in Section~\ref{3.3} suggest there is a chance that TOI-2257 b is a habitable planet with surface water. As is typical for planets orbiting M dwarfs with modern constraints, the biggest issue appears to be the loss of water during the pre-main sequence, when the XUV luminosity is high and the planet could be heated into a Venus-like state \citep{Luger&Barnes2015}. In this case, it appears that planet is likely to lose 10 TO during the pre-main sequence phase, which is less than for TRAPPIST-1 e \citep{Lincowski2018,Liuetal_2023,Giallucaetal_2024}, but still 1-2$\times$ more than all the water on Earth. This water loss is accompanied by a significant accumulation of abiotic oxygen, which can reach pressures of up to $\sim 15000$ bar in some simulations. Similar water-loss and oxygen-accumulation trends have been predicted for terrestrial planets around M dwarfs, with previous studies finding oxygen inventories of hundreds to thousands of bars following multi-TO water loss \citep{Luger&Barnes2015, Giallucaetal_2024}, although the exact magnitude depends strongly on the efficiency of oxygen sink.  While we do account for oxygen drag \citep{Hunten1987,Zahnle1990,Luger&Barnes2015}, we do not consider other oxygen sinks and thus our modeling represents an upper bound on oxygen accumulation. 

The eccentricity declines due to tidal dissipation, perhaps by 0.1 over the planet's expected lifetime if the initial eccentricity was larger than 0.6. Our simulations also show that this eccentricity evolution can be accompanied by inward migration, particularly for highly eccentric initial orbits. However, such large initial eccentricities are disfavored under the super-Earth scenario, since tidal damping would be expected to have substantially reduced the planet’s eccentricity by the present day. Thus, our results suggest that present-day eccentricities above $\sim0.62$ is consistent with the tidal evolution of a planet with gaseous composition. For the intermediate eccentricities more consistent with the super-Earth scenario, the change in semimajor axis is comparatively small. Thus, if TOI-2257 b is a super-Earth and has undergone tidal migration, our results suggest that it likely originated only modestly farther from the host star than its present-day orbit, rather than migrating inward from a substantially larger semimajor axis. In addition to driving orbital evolution, the large eccentricity should result in surface energy fluxes close to or exceeding Io, but whether this heating is strong enough to sterilize the planet is uncertain, see \citep{Barnesetal_2010}. Thus, our results indicate that tidal heating is not likely to rule out habitable surface conditions.

Tidal effects likely affect the planet in other significant ways than orbital circularization and the accompanying heating. The planet is almost assuredly tidally locked \citep{Barnes2017}, but probably not synchronously rotating \citep{Hut1981,Rodriguez2012}. Instead the planet is likely in a 2:1, or even faster rotating state depending on its eccentricity. The obliquity is likely damped to 0 \citep{Heller2011}, barring the presence of other companions \citep{Dobrovolskis2009,Correia2015,Guerrero2024}, so seasons would likely only be driven by instellation changes due to eccentricity. In other words, the seasons are global with summer at periastron and winter at apastron. Whether a planet on such a large eccentricity orbit is habitable remains to be seen, but some climate modeling work has suggested that they can be \citep{WilliamsPollard2002, Jietal_2023, Wayetal_2023, Liutal_2024}. Thus, at this time, we cannot rule out the possibility that TOI-2257 b is a habitable world.

\subsection{Outlook}

Our study is a first peek into the limits of a star's capacity to affect a highly eccentric transiting planet's eccentricity and atmospheric loss. Recent studies have demonstrated that stellar–planet interactions, such as tidal evolution, photoevaporation, and magnetic interactions, can shape the observed population of rocky planets \citep{Lee&Owen2025}. We caution that our simple models of stellar and planetary evolution are best for identifying trends, categories, and outliers, not predicting the current state of a specific planet. In this case, inspired by an actual target, we have established a baseline set of expectations for future research into the population of highly eccentric transiting planet. In this subsection we describe the most important simplifications in our model and suggest future research that could provide more insight into mini-Neptunes and super-Earths.

Probably the most important missing piece is the role of coronal mass ejections (CMEs) because almost no observations constrain their behavior for any star except the Sun \citep{Callingham2025}. Some guidance can be drawn from flares, as their sources appear highly correlated in the Sun \citep{Belov2007,Cliver2012,Youngblood2017}. The current flaring rate of TOI-2257 is unknown, but we can make some general statements based on its M3 spectral type and nominal 8 Gyr age. A reasonable comparison is GJ 1132 b, a lower mass M5 star with a similar age, but well-constrained XUV and flaring histories \citep{Barnesetal_2026}. They find that only 3 flares in 4 sectors of TESS data, which is barely sufficient to build a flare frequency distribution. In other words, that star is relatively quiet. This result is consistent with \citet{West2008} who mined the Sloan Digital Sky Survey (SDSS) data to show that only a few percent of thick disk M3 stars are active today. We may thus assume that the star is currently relatively inactive today.

In the past, however, flares and CMEs could have been much more powerful and the cumulative effect could have been profound. Figure~\ref{fig:4} shows that envelope masses larger than 0.2 M$_\oplus$ can be eroded quickly by thermal escape, so presumably are also susceptible to non-thermal escape driven by stellar energetic particles launched by CMEs. Should TOI-2257 b possess an envelope smaller than 0.02 M$_\oplus$, then that could be evidence for non-thermal escape.

A lower activity level today would improve the likelihood of acquiring high signal-to-noise JWST data during transit and/or eclipse. For example, measuring the orbital eccentricity via transit/eclipse timing asymmetries, acceleration during ingress and egress, or even refining the minimum eccentricity via the transit duration \citep{Barnes2007} would disfavor a terrestrial composition if the eccentricity is larger than 0.62, see Figure~\ref{fig:5}. A precise measurement of the transit and eclipse properties now could also provide a baseline for future observations that could search for changes. The high eccentricity also invites the possibility of a companion driving large amplitude eccentricity and/or inclination cycles \citep{Ford2000}, whose existence could be inferred if a precise measurement is taken in the near future and in several years to search for variations in transit and/or eclipse properties \citep{Agol2005,HolmanMurray2005}. 

Future theoretical work could add more physics to our model and/or higher resolution approaches. The contribution of flares to the total XUV budget could be included, \citep{Franceetal_2020,doAmaral_2022}. However, \cite{doAmaral_2022} found that XUV from flares remove less water as planetary mass increases. They simulated up to 5 Earth-masses and found that flares increase mass loss by about 10\% in that case, suggesting their role for this super-Earth version of TOI-2257 would be much less significant. The magnetohydrodynamics of the stellar wind interaction with the planetary atmosphere could also provide more insight into non-thermal mass loss mechanisms \citep{Cohen2015,Brain2016,Dong2018,Airapetian2020}.

Regarding tidal evolution, more realistic rheologies could be considered for the super-Earth case \citep[e.g.][]{Henning2009,Driscoll&Barnes_2015} in which the tidal dissipation is a function of mantle temperature. Such a model would be appropriate during the pre-HZ epoch of the planet when there are likely to be no oceans and would likely predict slower evolution \citep{Driscoll&Barnes_2015,Barnes2025lp890}. Should the planet form oceans, however, the tidal dissipation becomes a complicated function of continental shapes, bathymetry, and frequency, potentially leading to 5 orders of magnitude uncertainty in the pace of tidal evolution \citep{Blackledge2020}. Future research could directly simulate the physical oceanography of this planet to assess the range and role of oceanic dissipation in this extreme cases \citep[see, e.g.,][]{Shi2025}.

Should TOI-2257's eccentricity turn out to be $\sim 0.5$, it may be the first member of a population of high-eccentricity mini-Neptunes that could provide insight into the processes that shape the radius valley. Expanding our parameter sweep to a wider range of systems, particularly stellar mass and orbital period, could predict the edge of the valley in eccentricity space. The growing data set for M dwarfs \citep{Wanderley_2025} suggests theory and observations are converging and that population synthesis models of M dwarf mini-Neptunes could soon be constrained by data.

\section{Conclusion}
    \label{5}

We presented an investigation of the tidal and atmospheric evolution of TOI-2257 b, with a particular focus on its orbital eccentricity and atmospheric properties. We found that the planet appears to lie near the boundary between orbits and hydrogen envelopes that evolve significantly and those that don't. Thus, TOI-2257 b could serve as a crucial anchor point for models of the so-called radius valley, especially as a function of eccentricity. 

The large uncertainties in parameters and formation history, however, makes any conclusions about TOI-2257 b tentative. The star's old age could indicate a relatively quiet space weather environment and thus \jwst observations could be fruitful and confirm/refute the hypotheses that the planet currently possesses a hydrogen envelope and has an eccentricity in excess of 0.62. We found that the latter case would also be evidence that the planet supports a large gaseous envelope.

If atmospheric escape is mainly a thermal process, then we find that the envelope mass is most likely to be 0.1 M$_\oplus$ and larger than 0.02 M$_\oplus$. Unless the planet formed with a large envelope ($>0.2$ M$_\oplus$), mass loss has been modest throughout its lifetime. We hypothesize that if an envelope mass smaller than 0.02 M$_\oplus$ could be detected, then that would indicate non-thermal escape processes have sculpted the planet.

If TOI-2257 b formed as a terrestrial super-Earth, its primordial water inventory may have been destroyed by XUV radiation. This process could be critical for its potential to be habitable as we find most cases lose about twice as much water as in present on modern Earth. We also examined tidal heating on this world and find that while it could be at Io-like levels, it is not strong enough to trigger a runaway greenhouse. We thus conclude that TOI-2257 b is a potential abode for life. 

Our study has shown that many possibilities exist for the current and historical properties of this planet. Given its unique properties and the results of this study, more research into this planet is warranted. Whether or not the planet is gaseous, rocky, or even habitable, TOI-2257 b offers a unique opportunity to learn about the galactic population of exoplanets, the formation of and evolution of mini-Neptunes, and maybe even the discovery of life on a super-Earth. 

\section{Acknowledgment}
The authors thank Lhevy D. Geraldo and the two anonymous reviewers for their insightful comments, which greatly improved the quality of this paper. R.B. acknowledges support from NASA grant No. 80NSSC18K0829. M.K.A.O. acknowledges support from DOST-SEI and DLSU ASTHRDP-NSC.

\section{Data Availability}
The data underlying this article, including the input files and output data used in the analysis, are publicly available on Zenodo at \dataset[doi:10.5281/zenodo.23297276]
{https://doi.org/10.5281/zenodo.23297276}.

\bibliography{toi2257.bib}

@article{Barnes20,
	title = {{VPLanet}: {The} {Virtual} {Planet} {Simulator}},
	volume = {132},
	issn = {0004-6280, 1538-3873},
	shorttitle = {{VPLanet}},
	url = {https://iopscience.iop.org/article/10.1088/1538-3873/ab3ce8},
	doi = {10.1088/1538-3873/ab3ce8},
	language = {en},
	number = {1008},
	urldate = {2020-05-16},
	journal = {Publications of the Astronomical Society of the Pacific},
	author = {Barnes, Rory and Luger, Rodrigo and Deitrick, Russell and Driscoll, Peter and Quinn, Thomas R. and Fleming, David P. and Smotherman, Hayden and McDonald, Diego V. and Wilhelm, Caitlyn and Garcia, Rodolfo and Barth, Patrick and Guyer, Benjamin and Meadows, Victoria S. and Bitz, Cecilia M. and Gupta, Pramod and Domagal-Goldman, Shawn D. and Armstrong, John},
	month = feb,
	year = {2020},
	pages = {024502},}

@article{Schanche_2022,
   title={TOI-2257 b: A highly eccentric long-period sub-Neptune transiting a nearby M dwarf},
   volume={657},
   ISSN={1432-0746},
   url={http://dx.doi.org/10.1051/0004-6361/202142280},
   DOI={10.1051/0004-6361/202142280},
   journal={Astronomy I\& Astrophysics},
   publisher={EDP Sciences},
   author={Schanche, N. and Pozuelos, F. J. and Günther, M. N. and Wells, R. D. and Burgasser, A. J. and Chinchilla, P. and Delrez, L. and Ducrot, E. and Garcia, L. J. and Gómez Maqueo Chew, Y. and Jofré, E. and Rackham, B. V. and Sebastian, D. and Stassun, K. G. and Stern, D. and Timmermans, M. and Barkaoui, K. and Belinski, A. and Benkhaldoun, Z. and Benz, W. and Bieryla, A. and Bouchy, F. and Burdanov, A. and Charbonneau, D. and Christiansen, J. L. and Collins, K. A. and Demory, B.-O. and Dévora-Pajares, M. and de Wit, J. and Dragomir, D. and Dransfield, G. and Furlan, E. and Ghachoui, M. and Gillon, M. and Gnilka, C. and Gómez-Muñoz, M. A. and Guerrero, N. and Harris, M. and Heng, K. and Henze, C. E. and Hesse, K. and Howell, S. B. and Jehin, E. and Jenkins, J. and Jensen, E. L. N. and Kunimoto, M. and Latham, D. W. and Lester, K. and McLeod, K. K. and Mireles, I. and Murray, C. A. and Niraula, P. and Pedersen, P. P. and Queloz, D. and Quintana, E. V. and Ricker, G. and Rudat, A. and Sabin, L. and Safonov, B. and Schroffenegger, U. and Scott, N. and Seager, S. and Strakhov, I. and Triaud, A. H. M. J. and Vanderspek, R. and Vezie, M. and Winn, J.},
   year={2022},
   month=jan, pages={A45} }

@ARTICLE{Renaud2021,
       author = {{Renaud}, Joe P. and {Henning}, Wade G. and {Saxena}, Prabal and {Neveu}, Marc and {Bagheri}, Amirhossein and {Mandell}, Avi and {Hurford}, Terry},
        title = "{Tidal Dissipation in Dual-body, Highly Eccentric, and Nonsynchronously Rotating Systems: Applications to Pluto-Charon and the Exoplanet TRAPPIST-1e}",
      journal = {\psj},
         year = 2021,
        month = feb,
       volume = {2},
       number = {1},
          eid = {4},
        pages = {4},
          doi = {10.3847/PSJ/abc0f3},
archivePrefix = {arXiv},
       eprint = {2010.11801},
 primaryClass = {astro-ph.EP},
       adsurl = {https://ui.adsabs.harvard.edu/abs/2021PSJ.....2....4R}
}

@ARTICLE{Fulton2017,
       author = {{Fulton}, Benjamin J. and {Petigura}, Erik A. and {Howard}, Andrew W. and {Isaacson}, Howard and {Marcy}, Geoffrey W. and {Cargile}, Phillip A. and {Hebb}, Leslie and {Weiss}, Lauren M. and {Johnson}, John Asher and {Morton}, Timothy D. and {Sinukoff}, Evan and {Crossfield}, Ian J.~M. and {Hirsch}, Lea A.},
        title = "{The California-Kepler Survey. III. A Gap in the Radius Distribution of Small Planets}",
      journal = {\aj},
         year = 2017,
        month = sep,
       volume = {154},
       number = {3},
          eid = {109},
        pages = {109},
          doi = {10.3847/1538-3881/aa80eb},
archivePrefix = {arXiv},
       eprint = {1703.10375},
 primaryClass = {astro-ph.EP},
       adsurl = {https://ui.adsabs.harvard.edu/abs/2017AJ....154..109F}
}

@ARTICLE{Shi2025,
       author = {{Shi}, Jiaru and {Yang}, Jun and {Abbot}, Dorian S. and {Liu}, Yonggang and {Kang}, Wanying and {Lin}, Yufeng},
        title = "{Ocean Tides on Asynchronously Rotating Planets Orbiting Low-mass Stars}",
      journal = {\apj},
         year = 2025,
        month = aug,
       volume = {989},
       number = {2},
          eid = {139},
        pages = {139},
          doi = {10.3847/1538-4357/adeb83},
archivePrefix = {arXiv},
       eprint = {2507.03850},
 primaryClass = {astro-ph.EP},
       adsurl = {https://ui.adsabs.harvard.edu/abs/2025ApJ...989..139S}
}

@ARTICLE{Barnes2009,
       author = {{Barnes}, Rory and {Jackson}, Brian and {Raymond}, Sean N. and {West}, Andrew A. and {Greenberg}, Richard},
        title = "{The HD 40307 Planetary System: Super-Earths or Mini-Neptunes?}",
      journal = {\apj},
         year = 2009,
        month = apr,
       volume = {695},
       number = {2},
        pages = {1006-1011},
          doi = {10.1088/0004-637X/695/2/1006},
archivePrefix = {arXiv},
       eprint = {0901.1698},
 primaryClass = {astro-ph.EP},
       adsurl = {https://ui.adsabs.harvard.edu/abs/2009ApJ...695.1006B}
}

@ARTICLE{Blackledge2020,
       author = {{Blackledge}, B.~W. and {Green}, J.~A.~M. and {Barnes}, R. and {Way}, M.~J.},
        title = "{Tides on Other Earths: Implications for Exoplanet and Palaeo-Tidal Simulations}",
      journal = {\grl},
         year = 2020,
        month = jun,
       volume = {47},
       number = {12},
          eid = {e85746},
        pages = {e85746},
          doi = {10.1029/2019GL085746},
       adsurl = {https://ui.adsabs.harvard.edu/abs/2020GeoRL..4785746B}
}

@article{Chen_2017,
doi = {10.3847/1538-4357/834/1/17},
url = {https://dx.doi.org/10.3847/1538-4357/834/1/17},
year = {2016},
month = {dec},
publisher = {The American Astronomical Society},
volume = {834},
number = {1},
pages = {17},
author = {Jingjing Chen and David Kipping},
title = {PROBABILISTIC FORECASTING OF THE MASSES AND RADII OF OTHER WORLDS},
journal = {The Astrophysical Journal}
}

@article{Zeng_2016,
doi = {10.3847/0004-637X/819/2/127},
url = {https://dx.doi.org/10.3847/0004-637X/819/2/127},
year = {2016},
month = {mar},
publisher = {The American Astronomical Society},
volume = {819},
number = {2},
pages = {127},
author = {Li Zeng and Dimitar D. Sasselov and Stein B. Jacobsen},
title = {MASS–RADIUS RELATION FOR ROCKY PLANETS BASED ON PREM},
journal = {The Astrophysical Journal}
}

@ARTICLE{Xie_2014,
       author = {{Xie}, Ji-Wei},
        title = "{Transit Timing Variation of Near-resonance Planetary Pairs. II. Confirmation of 30 Planets in 15 Multiple-planet Systems}",
      journal = {\apjs},
         year = 2014,
        month = feb,
       volume = {210},
       number = {2},
          eid = {25},
        pages = {25},
          doi = {10.1088/0067-0049/210/2/25},
archivePrefix = {arXiv},
       eprint = {1309.2329},
 primaryClass = {astro-ph.EP},
       adsurl = {https://ui.adsabs.harvard.edu/abs/2014ApJS..210...25X}
}

@ARTICLE{Espinoza_2016,
       author = {{Espinoza}, N{\'e}stor and {Brahm}, Rafael and {Jord{\'a}n}, Andr{\'e}s and {Jenkins}, James S. and {Rojas}, Felipe and {Jofr{\'e}}, Paula and {M{\"a}dler}, Thomas and {Rabus}, Markus and {Chanam{\'e}}, Julio and {Pantoja}, Blake and {Soto}, Maritza G. and {Morzinski}, Katie M. and {Males}, Jared R. and {Ward-Duong}, Kimberly and {Close}, Laird M.},
        title = "{Discovery and Validation of a High-Density sub-Neptune from the K2 Mission}",
      journal = {\apj},
         year = 2016,
        month = oct,
       volume = {830},
       number = {1},
          eid = {43},
        pages = {43},
          doi = {10.3847/0004-637X/830/1/43},
archivePrefix = {arXiv},
       eprint = {1601.07608},
 primaryClass = {astro-ph.EP},
       adsurl = {https://ui.adsabs.harvard.edu/abs/2016ApJ...830...43E}
}

@ARTICLE{Wanderley_2025,
       author = {{Wanderley}, F{\'a}bio and {Cunha}, Katia and {Smith}, Verne V. and {Souto}, Diogo and {Pascucci}, Ilaria and {Behmard}, Aida and {Allende Prieto}, Carlos and {Beaton}, Rachael L. and {Bizyaev}, Dmitry and {Daflon}, Simone and {Hasselquist}, Sten and {Howell}, Steve and {Majewski}, Steven R. and {Pinsonneault}, Marc},
        title = "{An Analysis of the Radius Gap in a Sample of Kepler, K2 and TESS exoplanets orbiting M Dwarf Stars}",
      journal = {arXiv e-prints},
         year = 2025,
        month = sep,
          eid = {arXiv:2509.01930},
        pages = {arXiv:2509.01930},
          doi = {10.48550/arXiv.2509.01930},
archivePrefix = {arXiv},
       eprint = {2509.01930},
 primaryClass = {astro-ph.EP},
       adsurl = {https://ui.adsabs.harvard.edu/abs/2025arXiv250901930W}
}

@article{Jackson_2012,
    author = {Jackson, Alan P. and Davis, Timothy A. and Wheatley, Peter J.},
    title = "{The coronal X-ray–age relation and its implications for the evaporation of exoplanets}",
    journal = {Monthly Notices of the Royal Astronomical Society},
    volume = {422},
    number = {3},
    pages = {2024-2043},
    year = {2012},
    month = {05},
    issn = {0035-8711},
    doi = {10.1111/j.1365-2966.2012.20657.x},
    url = {https://doi.org/10.1111/j.1365-2966.2012.20657.x},
    eprint = {https://academic.oup.com/mnras/article-pdf/422/3/2024/18749111/mnras0422-2024.pdf},
}

@article{ Baraffe,
	author = {Baraffe, Isabelle and Homeier, Derek and Allard, France and Chabrier, Gilles},
	title = {New evolutionary models for pre-main sequence and main sequence low-mass stars down to the hydrogen-burning limit },
	DOI= "10.1051/0004-6361/201425481",
	url= "https://doi.org/10.1051/0004-6361/201425481",
	journal = {A\&A},
	year = 2015,
	volume = 577,
	pages = "A42",
	month = "",
}

@article{Ribas_2005,
doi = {10.1086/427977},
url = {https://dx.doi.org/10.1086/427977},
year = {2005},
month = {mar},
publisher = {},
volume = {622},
number = {1},
pages = {680},
author = {Ignasi Ribas and Edward F. Guinan and Manuel Güdel and Marc Audard},
title = {Evolution of the Solar Activity over Time and Effects on Planetary Atmospheres. I. High-Energy Irradiances (1-1700 Å)},
journal = {The Astrophysical Journal}
}

@article{Jackson_2008,
doi = {10.1086/529187},
url = {https://dx.doi.org/10.1086/529187},
year = {2008},
month = {may},
publisher = {},
volume = {678},
number = {2},
pages = {1396},
author = {Brian Jackson and Richard Greenberg and Rory Barnes},
title = {Tidal Evolution of Close-in Extrasolar Planets},
journal = {The Astrophysical Journal}
}

@article{Barnes_2008,
   title={Tides and the Evolution of Planetary Habitability},
   volume={8},
   ISSN={1557-8070},
   url={http://dx.doi.org/10.1089/ast.2007.0204},
   DOI={10.1089/ast.2007.0204},
   number={3},
   journal={Astrobiology},
   publisher={Mary Ann Liebert Inc},
   author={Barnes, Rory and Raymond, Sean N. and Jackson, Brian and Greenberg, Richard},
   year={2008},
   month=jun, pages={557–568} }

@misc{Penev_2024,
      title={Main-sequence exoplanet systems: tidal evolution}, 
      author={Kaloyan Penev},
      year={2024},
      eprint={2403.07716},
      archivePrefix={arXiv},
      primaryClass={astro-ph.EP},
      url={https://arxiv.org/abs/2403.07716}, 
}

@article{Darwin_1880,
author = {Darwin, George Howard },
title = {XX. On the secular changes in the elements of the orbit of a satellite revolving about a tidally distorted planet},
journal = {Philosophical Transactions of the Royal Society of London},
volume = {171},
number = {},
pages = {713-891},
year = {1880},
doi = {10.1098/rstl.1880.0020},

URL = {https://royalsocietypublishing.org/doi/abs/10.1098/rstl.1880.0020},
eprint = {https://royalsocietypublishing.org/doi/pdf/10.1098/rstl.1880.0020}
}

@article{Rogers_2011,
doi = {10.1088/0004-637X/738/1/59},
url = {https://dx.doi.org/10.1088/0004-637X/738/1/59},
year = {2011},
month = {aug},
publisher = {The American Astronomical Society},
volume = {738},
number = {1},
pages = {59},
author = {Leslie A. Rogers and Peter Bodenheimer and Jack J. Lissauer and Sara Seager},
title = {FORMATION AND STRUCTURE OF LOW-DENSITY EXO-NEPTUNES},
journal = {The Astrophysical Journal}
}

@ARTICLE{Rogers2015,
       author = {{Rogers}, Leslie A.},
        title = "{Most 1.6 Earth-radius Planets are Not Rocky}",
      journal = {\apj},
         year = 2015,
        month = mar,
       volume = {801},
       number = {1},
          eid = {41},
        pages = {41},
          doi = {10.1088/0004-637X/801/1/41},
archivePrefix = {arXiv},
       eprint = {1407.4457},
 primaryClass = {astro-ph.EP},
       adsurl = {https://ui.adsabs.harvard.edu/abs/2015ApJ...801...41R}
}

@ARTICLE{Belov2007,
       author = {{Belov}, A. and {Kurt}, V. and {Mavromichalaki}, H. and {Gerontidou}, M.},
        title = "{Peak-Size Distributions of Proton Fluxes and Associated Soft X-Ray Flares}",
      journal = {\solphys},
         year = 2007,
        month = dec,
       volume = {246},
       number = {2},
        pages = {457-470},
          doi = {10.1007/s11207-007-9071-x},
       adsurl = {https://ui.adsabs.harvard.edu/abs/2007SoPh..246..457B}
}

@ARTICLE{Cliver2012,
       author = {{Cliver}, E.~W. and {Ling}, A.~G. and {Belov}, A. and {Yashiro}, S.},
        title = "{Size distributions of solar flares and solar energetic particle events}",
      journal = {\apjl},
         year = 2012,
        month = sep,
       volume = {756},
       number = {2},
          eid = {L29},
        pages = {L29},
          doi = {10.1088/2041-8205/756/2/L29},
       adsurl = {https://ui.adsabs.harvard.edu/abs/2012ApJ...756L..29C}
}

@ARTICLE{Callingham2025,
       author = {{Callingham}, J.~R. and {Tasse}, C. and {Keers}, R. and {Kavanagh}, R.~D. and {Vedantham}, H. and {Zarka}, P. and {Bellotti}, S. and {Cristofari}, P.~I. and {Bloot}, S. and {Konijn}, D.~C. and {Hardcastle}, M.~J. and {Lamy}, L. and {Pass}, E.~K. and {Pope}, B.~J.~S. and {Reid}, H. and {R{\"o}ttgering}, H.~J.~A. and {Shimwell}, T.~W. and {Zucca}, P.},
        title = "{Radio Burst from a Stellar Coronal Mass Ejection}",
      journal = {Nature},
         year = 2025,
        month = nov,
          eid = {arXiv:2511.09289},
        pages = {arXiv:2511.09289},
          doi = {10.48550/arXiv.2511.09289},
archivePrefix = {arXiv},
       eprint = {2511.09289},
 primaryClass = {astro-ph.SR},
       adsurl = {https://ui.adsabs.harvard.edu/abs/2025arXiv251109289C}
}

@ARTICLE{Youngblood2017,
       author = {{Youngblood}, Allison and {France}, Kevin and {Loyd}, R.~O. Parke and {Brown}, Alexander and {Mason}, James P. and {Schneider}, P. Christian and {Tilley}, Matt A. and {Berta-Thompson}, Zachory K. and {Buccino}, Andrea and {Froning}, Cynthia S. and {Hawley}, Suzanne L. and {Linsky}, Jeffrey and {Mauas}, Pablo J.~D. and {Redfield}, Seth and {Kowalski}, Adam and {Miguel}, Yamila and {Newton}, Elisabeth R. and {Rugheimer}, Sarah and {Segura}, Ant{\'\i}gona and {Roberge}, Aki and {Vieytes}, Mariela},
        title = "{The MUSCLES Treasury Survey. IV. Scaling Relations for Ultraviolet, Ca II K, and Energetic Particle Fluxes from M Dwarfs}",
      journal = {\apj},
         year = 2017,
        month = jul,
       volume = {843},
       number = {1},
          eid = {31},
        pages = {31},
          doi = {10.3847/1538-4357/aa76dd},
archivePrefix = {arXiv},
       eprint = {1705.04361},
 primaryClass = {astro-ph.SR},
       adsurl = {https://ui.adsabs.harvard.edu/abs/2017ApJ...843...31Y}
}

@ARTICLE{Raymond2008,
       author = {{Raymond}, Sean N. and {Barnes}, Rory and {Mandell}, Avi M.},
        title = "{Observable consequences of planet formation models in systems with close-in terrestrial planets}",
      journal = {\mnras},
         year = 2008,
        month = feb,
       volume = {384},
       number = {2},
        pages = {663-674},
          doi = {10.1111/j.1365-2966.2007.12712.x},
archivePrefix = {arXiv},
       eprint = {0711.2015},
 primaryClass = {astro-ph},
       adsurl = {https://ui.adsabs.harvard.edu/abs/2008MNRAS.384..663R}
}

@ARTICLE{jackson2010,
       author = {{Jackson}, Brian and {Miller}, Neil and {Barnes}, Rory and {Raymond}, Sean N. and {Fortney}, Jonathan J. and {Greenberg}, Richard},
        title = "{The roles of tidal evolution and evaporative mass loss in the origin of CoRoT-7 b}",
      journal = {\mnras},
         year = 2010,
        month = sep,
       volume = {407},
       number = {2},
        pages = {910-922},
          doi = {10.1111/j.1365-2966.2010.17012.x},
archivePrefix = {arXiv},
       eprint = {1005.2186},
 primaryClass = {astro-ph.EP},
       adsurl = {https://ui.adsabs.harvard.edu/abs/2010MNRAS.407..910J}
}

@ARTICLE{Loyd2025,
       author = {{Loyd}, R.~O. Parke and {Schreyer}, Ethan and {Owen}, James E. and {Rogers}, James G. and {Broome}, Madelyn I. and {Shkolnik}, Evgenya L. and {Murray-Clay}, Ruth and {Wilson}, David J. and {Peacock}, Sarah and {Teske}, Johanna and {Schlichting}, Hilke E. and {Duvvuri}, Girish M. and {Youngblood}, Allison and {Schneider}, P. Christian and {France}, Kevin and {Giacalone}, Steven and {Batalha}, Natasha E. and {Schneider}, Adam C. and {Longo}, Isabella and {Barman}, Travis and {Ardila}, David R.},
        title = "{Hydrogen escaping from a pair of exoplanets smaller than Neptune}",
      journal = {\nat},
         year = 2025,
        month = feb,
       volume = {638},
       number = {8051},
        pages = {636-639},
          doi = {10.1038/s41586-024-08490-x},
       adsurl = {https://ui.adsabs.harvard.edu/abs/2025Natur.638..636L}
}

@ARTICLE{Kasting1993,
       author = {{Kasting}, James F. and {Whitmire}, Daniel P. and {Reynolds}, Ray T.},
        title = "{Habitable Zones around Main Sequence Stars}",
      journal = {\icarus},
         year = 1993,
        month = jan,
       volume = {101},
       number = {1},
        pages = {108-128},
          doi = {10.1006/icar.1993.1010},
       adsurl = {https://ui.adsabs.harvard.edu/abs/1993Icar..101..108K}
}

@article{WATSON1981150,
title = {The dynamics of a rapidly escaping atmosphere: Applications to the evolution of Earth and Venus},
journal = {Icarus},
volume = {48},
number = {2},
pages = {150-166},
year = {1981},
issn = {0019-1035},
doi = {https://doi.org/10.1016/0019-1035(81)90101-9},
url = {https://www.sciencedirect.com/science/article/pii/0019103581901019},
author = {Andrew J. Watson and Thomas M. Donahue and James C.G. Walker}
}

@article{Erkaev_2007,
   title={Roche lobe effects on the atmospheric  loss
from “Hot Jupiters”},
   volume={472},
   ISSN={1432-0746},
   url={http://dx.doi.org/10.1051/0004-6361:20066929},
   DOI={10.1051/0004-6361:20066929},
   number={1},
   journal={Astronomy I\& Astrophysics},
   publisher={EDP Sciences},
   author={Erkaev, N. V. and Kulikov, Yu. N. and Lammer, H. and Selsis, F. and Langmayr, D. and Jaritz, G. F. and Biernat, H. K.},
   year={2007},
   month=jul, pages={329–334} }

@article{Lammer,
    author = {Lammer, H. and Erkaev, N. V. and Odert, P. and Kislyakova, K. G. and Leitzinger, M. and Khodachenko, M. L.},
    title = "{Probing the blow-off criteria of hydrogen-rich ‘super-Earths’}",
    journal = {Monthly Notices of the Royal Astronomical Society},
    volume = {430},
    number = {2},
    pages = {1247-1256},
    year = {2013},
    month = {01},
    issn = {0035-8711},
    doi = {10.1093/mnras/sts705},
    url = {https://doi.org/10.1093/mnras/sts705},
    eprint = {https://academic.oup.com/mnras/article-pdf/430/2/1247/9382479/sts705.pdf},
}

@article{Volkov_2013,
doi = {10.1088/0004-637X/765/2/90},
url = {https://dx.doi.org/10.1088/0004-637X/765/2/90},
year = {2013},
month = {feb},
publisher = {The American Astronomical Society},
volume = {765},
number = {2},
pages = {90},
author = {Alexey N. Volkov and Robert E. Johnson},
title = {THERMAL ESCAPE IN THE HYDRODYNAMIC REGIME: RECONSIDERATION OF PARKER's ISENTROPIC THEORY BASED ON RESULTS OF KINETIC SIMULATIONS},
journal = {The Astrophysical Journal}
}

@article{Johnson_2013,
doi = {10.1088/2041-8205/768/1/L4},
url = {https://dx.doi.org/10.1088/2041-8205/768/1/L4},
year = {2013},
month = {apr},
publisher = {The American Astronomical Society},
volume = {768},
number = {1},
pages = {L4},
author = {Robert E. Johnson and Alexey N. Volkov and Justin T. Erwin},
title = {MOLECULAR-KINETIC SIMULATIONS OF ESCAPE FROM THE EX-PLANET AND EXOPLANETS: CRITERION FOR TRANSONIC FLOW},
journal = {The Astrophysical Journal Letters}
}

@article{Murray-Clay_2009,
doi = {10.1088/0004-637X/693/1/23},
url = {https://dx.doi.org/10.1088/0004-637X/693/1/23},
year = {2009},
month = {feb},
publisher = {The American Astronomical Society},
volume = {693},
number = {1},
pages = {23},
author = {Ruth A. Murray-Clay and Eugene I. Chiang and Norman Murray},
title = {ATMOSPHERIC ESCAPE FROM HOT JUPITERS},
journal = {The Astrophysical Journal}
}

@article{OwenJackson,
    author = {Owen, James E. and Jackson, Alan P.},
    title = "{Planetary evaporation by UV and X-ray radiation: basic hydrodynamics}",
    journal = {Monthly Notices of the Royal Astronomical Society},
    volume = {425},
    number = {4},
    pages = {2931-2947},
    year = {2012},
    month = {10},
    issn = {0035-8711},
    doi = {10.1111/j.1365-2966.2012.21481.x},
    url = {https://doi.org/10.1111/j.1365-2966.2012.21481.x},
    eprint = {https://academic.oup.com/mnras/article-pdf/425/4/2931/4909170/425-4-2931.pdf},
}

@article{Owen_2016,
doi = {10.3847/0004-637X/816/1/34},
url = {https://dx.doi.org/10.3847/0004-637X/816/1/34},
year = {2015},
month = {dec},
publisher = {The American Astronomical Society},
volume = {816},
number = {1},
pages = {34},
author = {James E. Owen and Marcelo A. Alvarez},
title = {UV DRIVEN EVAPORATION OF CLOSE-IN PLANETS: ENERGY-LIMITED, RECOMBINATION-LIMITED, AND PHOTON-LIMITED FLOWS},
journal = {The Astrophysical Journal}
}

@article{luger2015habitable,
  title={Habitable evaporated cores: transforming mini-Neptunes into super-Earths in the habitable zones of M dwarfs},
  author={Luger, Rodrigo and Barnes, Rory and Lopez, Eric and Fortney, Jonathan and Jackson, Brian and Meadows, Victoria},
  journal={Astrobiology},
  doi={10.1089/ast.2014.1215},
  volume={15},
  number={1},
  pages={57--88},
  year={2015},
  publisher={Mary Ann Liebert, Inc. 140 Huguenot Street, 3rd Floor New Rochelle, NY 10801 USA}
}

@article{owen2016atmospheres,
  title={Atmospheres of low-mass planets: the “boil-off”},
  author={Owen, James E and Wu, Yanqin},
  journal={The Astrophysical Journal},
  volume={817},
  number={2},
  pages={107},
  year={2016},
  publisher={IOP Publishing}
}

@article{doAmaral_2022,
doi = {10.3847/1538-4357/ac53af},
url = {https://dx.doi.org/10.3847/1538-4357/ac53af},
year = {2022},
month = {mar},
publisher = {The American Astronomical Society},
volume = {928},
number = {1},
pages = {12},
author = {Laura N. R. do Amaral and Rory Barnes and Antígona Segura and Rodrigo Luger},
title = {The Contribution of M-dwarf Flares to the Thermal Escape of Potentially Habitable Planet Atmospheres},
journal = {The Astrophysical Journal}
}

@article{Gupta,
    author = {Gupta, Akash and Nicholson, Lorraine and Schlichting, Hilke E},
    title = "{Properties of the radius valley around low mass stars: predictions from the core-powered mass-loss mechanism}",
    journal = {Monthly Notices of the Royal Astronomical Society},
    volume = {516},
    number = {3},
    pages = {4585-4593},
    year = {2022},
    month = {09},
    issn = {0035-8711},
    doi = {10.1093/mnras/stac2488},
    url = {https://doi.org/10.1093/mnras/stac2488},
    eprint = {https://academic.oup.com/mnras/article-pdf/516/3/4585/45985493/stac2488.pdf},
}

@article{leconte2010tidal,
  title={Is tidal heating sufficient to explain bloated exoplanets? Consistent calculations accounting for finite initial eccentricity},
  author={Leconte, J{\'e}r{\'e}my and Chabrier, Gilles and Baraffe, Isabelle and Levrard, Benjamin},
  journal={Astronomy \& Astrophysics},
  doi={10.1051/0004-6361/201014337},
  url={https://doi.org/10.1051/0004-6361/201014337},
  volume={516},
  pages={A64},
  year={2010},
  publisher={EDP Sciences}
}

@article{ferraz2008tidal,
  title={Tidal friction in close-in satellites and exoplanets: The Darwin theory re-visited},
  author={Ferraz-Mello, Sylvio and Rodr{\'\i}guez, Adri{\'a}n and Hussmann, Hauke},
  journal={Celestial Mechanics and Dynamical Astronomy},
  doi={10.1007/s10569-008-9133-x},
  url={https://doi.org/10.1007/s10569-008-9133-x},
  volume={101},
  pages={171--201},
  year={2008},
  publisher={Springer}
}

@article{Luger&Barnes2015,
  title={Extreme Water Loss and Abiotic O2 Buildup on Planets Throughout the Habitable Zones of M Dwarfs},
  author={Rodrigo Luger and Rory Barnes},
  DOI= "10.1089/ast.2014.1231",
  url= "https://doi.org/10.1089/ast.2014.1231",
  journal={Astrobiology},
  volume={15},
  pages={119–143},
  year={2015},
  publisher={Mary Ann Liebert, Inc.}
}

@ARTICLE{Williams1997,
       author = {{Williams}, Darren M. and {Kasting}, James F. and {Wade}, Richard A.},
        title = "{Habitable moons around extrasolar giant planets}",
      journal = {\nat},
         year = 1997,
        month = jan,
       volume = {385},
       number = {6613},
        pages = {234-236},
          doi = {10.1038/385234a0},
       adsurl = {https://ui.adsabs.harvard.edu/abs/1997Natur.385..234W}
}

@ARTICLE{HellerBarnes2013,
       author = {{Heller}, Ren{\'e} and {Barnes}, Rory},
        title = "{Exomoon Habitability Constrained by Illumination and Tidal Heating}",
      journal = {Astrobiology},
         year = 2013,
        month = jan,
       volume = {13},
       number = {1},
        pages = {18-46},
          doi = {10.1089/ast.2012.0859},
archivePrefix = {arXiv},
       eprint = {1209.5323},
 primaryClass = {astro-ph.EP},
       adsurl = {https://ui.adsabs.harvard.edu/abs/2013AsBio..13...18H}
}

@ARTICLE{Heller2014,
       author = {{Heller}, Ren{\'e} and {Williams}, Darren and {Kipping}, David and {Limbach}, Mary Anne and {Turner}, Edwin and {Greenberg}, Richard and {Sasaki}, Takanori and {Bolmont}, {\'E}meline and {Grasset}, Olivier and {Lewis}, Karen and {Barnes}, Rory and {Zuluaga}, Jorge I.},
        title = "{Formation, Habitability, and Detection of Extrasolar Moons}",
      journal = {Astrobiology},
         year = 2014,
        month = sep,
       volume = {14},
       number = {9},
        pages = {798-835},
          doi = {10.1089/ast.2014.1147},
archivePrefix = {arXiv},
       eprint = {1408.6164},
 primaryClass = {astro-ph.EP},
       adsurl = {https://ui.adsabs.harvard.edu/abs/2014AsBio..14..798H}
}

@ARTICLE{Hayashi1961,
       author = {{Hayashi}, Chushiro},
        title = "{Stellar evolution in early phases of gravitational contraction.}",
      journal = {\pasj},
         year = 1961,
        month = dec,
       volume = {13},
        pages = {450-452},
       adsurl = {https://ui.adsabs.harvard.edu/abs/1961PASJ...13..450H}
}

@ARTICLE{Lambrechts19,
       author = {{Lambrechts}, Michiel and {Morbidelli}, Alessandro and {Jacobson}, Seth A. and {Johansen}, Anders and {Bitsch}, Bertram and {Izidoro}, Andre and {Raymond}, Sean N.},
        title = "{Formation of planetary systems by pebble accretion and migration. How the radial pebble flux determines a terrestrial-planet or super-Earth growth mode}",
      journal = {\aap},
         year = 2019,
        month = jul,
       volume = {627},
          eid = {A83},
        pages = {A83},
          doi = {10.1051/0004-6361/201834229},
archivePrefix = {arXiv},
       eprint = {1902.08694},
 primaryClass = {astro-ph.EP},
       adsurl = {https://ui.adsabs.harvard.edu/abs/2019A&A...627A..83L}
}

@ARTICLE{Clement22,
       author = {{Clement}, Matthew S. and {Quintana}, Elisa V. and {Quarles}, Billy L.},
        title = "{Habitable Planet Formation around Low-mass Stars: Rapid Accretion, Rapid Debris Removal, and the Essential Contribution of External Giants}",
      journal = {\apj},
         year = 2022,
        month = mar,
       volume = {928},
       number = {1},
          eid = {91},
        pages = {91},
          doi = {10.3847/1538-4357/ac549e},
archivePrefix = {arXiv},
       eprint = {2202.05909},
 primaryClass = {astro-ph.EP},
       adsurl = {https://ui.adsabs.harvard.edu/abs/2022ApJ...928...91C}
}

@ARTICLE{Raymond2007,
       author = {{Raymond}, Sean N. and {Scalo}, John and {Meadows}, Victoria S.},
        title = "{A Decreased Probability of Habitable Planet Formation around Low-Mass Stars}",
      journal = {\apj},
         year = 2007,
        month = nov,
       volume = {669},
       number = {1},
        pages = {606-614},
          doi = {10.1086/521587},
archivePrefix = {arXiv},
       eprint = {0707.1711},
 primaryClass = {astro-ph},
       adsurl = {https://ui.adsabs.harvard.edu/abs/2007ApJ...669..606R}
}

@ARTICLE{Rasio1996,
       author = {{Rasio}, F.~A. and {Tout}, C.~A. and {Lubow}, S.~H. and {Livio}, M.},
        title = "{Tidal Decay of Close Planetary Orbits}",
      journal = {\apj},
         year = 1996,
        month = oct,
       volume = {470},
        pages = {1187},
          doi = {10.1086/177941},
archivePrefix = {arXiv},
       eprint = {astro-ph/9605059},
 primaryClass = {astro-ph},
       adsurl = {https://ui.adsabs.harvard.edu/abs/1996ApJ...470.1187R}
}

@ARTICLE{Levrard2007,
       author = {{Levrard}, B. and {Correia}, A.~C.~M. and {Chabrier}, G. and {Baraffe}, I. and {Selsis}, F. and {Laskar}, J.},
        title = "{Tidal dissipation within hot Jupiters: a new appraisal}",
      journal = {\aap},
         year = 2007,
        month = jan,
       volume = {462},
       number = {1},
        pages = {L5-L8},
          doi = {10.1051/0004-6361:20066487},
archivePrefix = {arXiv},
       eprint = {astro-ph/0612044},
 primaryClass = {astro-ph},
       adsurl = {https://ui.adsabs.harvard.edu/abs/2007A&A...462L...5L}
}

@ARTICLE{Barnes2017,
       author = {{Barnes}, Rory},
        title = "{Tidal locking of habitable exoplanets}",
      journal = {Celestial Mechanics and Dynamical Astronomy},
         year = 2017,
        month = dec,
       volume = {129},
       number = {4},
        pages = {509-536},
          doi = {10.1007/s10569-017-9783-7},
archivePrefix = {arXiv},
       eprint = {1708.02981},
 primaryClass = {astro-ph.EP},
       adsurl = {https://ui.adsabs.harvard.edu/abs/2017CeMDA.129..509B}
}

@ARTICLE{Jackson2008habitable,
       author = {{Jackson}, Brian and {Barnes}, Rory and {Greenberg}, Richard},
        title = "{Tidal heating of terrestrial extrasolar planets and implications for their habitability}",
      journal = {\mnras},
         year = 2008,
        month = nov,
       volume = {391},
       number = {1},
        pages = {237-245},
          doi = {10.1111/j.1365-2966.2008.13868.x},
archivePrefix = {arXiv},
       eprint = {0808.2770},
 primaryClass = {astro-ph},
       adsurl = {https://ui.adsabs.harvard.edu/abs/2008MNRAS.391..237J}
}

@ARTICLE{West2008,
       author = {{West}, Andrew A. and {Hawley}, Suzanne L. and {Bochanski}, John J. and {Covey}, Kevin R. and {Reid}, I. Neill and {Dhital}, Saurav and {Hilton}, Eric J. and {Masuda}, Michael},
        title = "{Constraining the Age-Activity Relation for Cool Stars: The Sloan Digital Sky Survey Data Release 5 Low-Mass Star Spectroscopic Sample}",
      journal = {\aj},
         year = 2008,
        month = mar,
       volume = {135},
       number = {3},
        pages = {785-795},
          doi = {10.1088/0004-6256/135/3/785},
archivePrefix = {arXiv},
       eprint = {0712.1590},
 primaryClass = {astro-ph},
       adsurl = {https://ui.adsabs.harvard.edu/abs/2008AJ....135..785W}
}

@ARTICLE{Engle2024,
       author = {{Engle}, Scott G.},
        title = "{Living with a Red Dwarf: X-Ray, UV, and Ca II Activity-Age Relationships of M Dwarfs}",
      journal = {\apj},
         year = 2024,
        month = jan,
       volume = {960},
       number = {1},
          eid = {62},
        pages = {62},
          doi = {10.3847/1538-4357/ad0840},
archivePrefix = {arXiv},
       eprint = {2310.04302},
 primaryClass = {astro-ph.SR},
       adsurl = {https://ui.adsabs.harvard.edu/abs/2024ApJ...960...62E}
}

@ARTICLE{GoldreichSoter1966,
       author = {{Goldreich}, Peter and {Soter}, Steven},
        title = "{Q in the Solar System}",
      journal = {\icarus},
         year = 1966,
        month = jan,
       volume = {5},
       number = {1},
        pages = {375-389},
          doi = {10.1016/0019-1035(66)90051-0},
       adsurl = {https://ui.adsabs.harvard.edu/abs/1966Icar....5..375G}
}

@ARTICLE{AksnesFranklin2001,
       author = {{Aksnes}, K. and {Franklin}, F.~A.},
        title = "{Secular Acceleration of Io Derived from Mutual Satellite Events}",
      journal = {\aj},
         year = 2001,
        month = nov,
       volume = {122},
       number = {5},
        pages = {2734-2739},
          doi = {10.1086/323708},
       adsurl = {https://ui.adsabs.harvard.edu/abs/2001AJ....122.2734A}
}

@ARTICLE{Henning2009,
       author = {{Henning}, Wade G. and {O'Connell}, Richard J. and {Sasselov}, Dimitar D.},
        title = "{Tidally Heated Terrestrial Exoplanets: Viscoelastic Response Models}",
      journal = {\apj},
         year = 2009,
        month = dec,
       volume = {707},
       number = {2},
        pages = {1000-1015},
          doi = {10.1088/0004-637X/707/2/1000},
archivePrefix = {arXiv},
       eprint = {0912.1907},
 primaryClass = {astro-ph.EP},
       adsurl = {https://ui.adsabs.harvard.edu/abs/2009ApJ...707.1000H}
}

@ARTICLE{Jackson2008heating,
       author = {{Jackson}, Brian and {Greenberg}, Richard and {Barnes}, Rory},
        title = "{Tidal Heating of Extrasolar Planets}",
      journal = {\apj},
         year = 2008,
        month = jul,
       volume = {681},
       number = {2},
        pages = {1631-1638},
          doi = {10.1086/587641},
archivePrefix = {arXiv},
       eprint = {0803.0026},
 primaryClass = {astro-ph},
       adsurl = {https://ui.adsabs.harvard.edu/abs/2008ApJ...681.1631J}
}

@ARTICLE{Barnes2010,
       author = {{Barnes}, Rory and {Raymond}, Sean N. and {Greenberg}, Richard and {Jackson}, Brian and {Kaib}, Nathan A.},
        title = "{CoRoT-7b: Super-Earth or Super-Io?}",
      journal = {\apjl},
         year = 2010,
        month = feb,
       volume = {709},
       number = {2},
        pages = {L95-L98},
          doi = {10.1088/2041-8205/709/2/L95},
archivePrefix = {arXiv},
       eprint = {0912.1337},
 primaryClass = {astro-ph.EP},
       adsurl = {https://ui.adsabs.harvard.edu/abs/2010ApJ...709L..95B}
}

@ARTICLE{Bolmont2017,
       author = {{Bolmont}, E. and {Selsis}, F. and {Owen}, J.~E. and {Ribas}, I. and {Raymond}, S.~N. and {Leconte}, J. and {Gillon}, M.},
        title = "{Water loss from terrestrial planets orbiting ultracool dwarfs: implications for the planets of TRAPPIST-1}",
      journal = {\mnras},
         year = 2017,
        month = jan,
       volume = {464},
       number = {3},
        pages = {3728-3741},
          doi = {10.1093/mnras/stw2578},
archivePrefix = {arXiv},
       eprint = {1605.00616},
 primaryClass = {astro-ph.EP},
       adsurl = {https://ui.adsabs.harvard.edu/abs/2017MNRAS.464.3728B}
}

@ARTICLE{Barth2021,
       author = {{Barth}, Patrick and {Carone}, Ludmila and {Barnes}, Rory and {Noack}, Lena and {Molli{\`e}re}, Paul and {Henning}, Thomas},
        title = "{Magma Ocean Evolution of the TRAPPIST-1 Planets}",
      journal = {Astrobiology},
         year = 2021,
        month = nov,
       volume = {21},
       number = {11},
        pages = {1325-1349},
          doi = {10.1089/ast.2020.2277},
archivePrefix = {arXiv},
       eprint = {2008.09599},
 primaryClass = {astro-ph.EP},
       adsurl = {https://ui.adsabs.harvard.edu/abs/2021AsBio..21.1325B}
}

@ARTICLE{Carone2025,
       author = {{Carone}, L. and {Barnes}, R. and {Noack}, L. and {Chubb}, K. and {Barth}, P. and {Bitsch}, B. and {Thamm}, A. and {Balduin}, A. and {Garcia}, R. and {Helling}, Ch.},
        title = "{From CO$_{2}$- to H$_{2}$O-dominated atmospheres and back: How mixed outgassing changes the volatile distribution in magma oceans around M dwarf stars}",
      journal = {\aap},
         year = 2025,
        month = jan,
       volume = {693},
          eid = {A303},
        pages = {A303},
          doi = {10.1051/0004-6361/202450307},
archivePrefix = {arXiv},
       eprint = {2412.10192},
 primaryClass = {astro-ph.EP},
       adsurl = {https://ui.adsabs.harvard.edu/abs/2025A&A...693A.303C}
}

@ARTICLE{Lambeck1977,
   author = {{Lambeck}, K.},
    title = "{Tidal Dissipation in the Oceans: Astronomical, Geophysical and Oceanographic Consequences}",
  journal = {Royal Society of London Philosophical Transactions Series A},
     year = 1977,
    month = dec,
   volume = 287,
    pages = {545-594},
      doi = {10.1098/rsta.1977.0159},
   adsurl = {http://adsabs.harvard.edu/abs/1977RSPTA.287..545L}
}

@ARTICLE{Veeder1994,
       author = {{Veeder}, Glenn J. and {Matson}, Dennis L. and {Johnson}, Torrence V. and {Blaney}, Diana L. and {Goguen}, Jay D.},
        title = "{Io's heat flow from infrared radiometry: 1983-1993}",
      journal = {\jgr},
         year = 1994,
        month = aug,
       volume = {99},
       number = {E8},
        pages = {17095-17162},
          doi = {10.1029/94JE00637},
       adsurl = {https://ui.adsabs.harvard.edu/abs/1994JGR....9917095V}
}

@ARTICLE{Abe1993,
       author = {{Abe}, Yutaka},
        title = "{Physical state of the very early Earth}",
      journal = {Lithos},
         year = 1993,
        month = sep,
       volume = {30},
       number = {3},
        pages = {223-235},
          doi = {10.1016/0024-4937(93)90037-D},
       adsurl = {https://ui.adsabs.harvard.edu/abs/1993Litho..30..223A}
}

@ARTICLE{Goldblatt2013,
       author = {{Goldblatt}, Colin and {Robinson}, Tyler D. and {Zahnle}, Kevin J. and {Crisp}, David},
        title = "{Low simulated radiation limit for runaway greenhouse climates}",
      journal = {Nature Geoscience},
         year = 2013,
        month = aug,
       volume = {6},
       number = {8},
        pages = {661-667},
          doi = {10.1038/ngeo1892},
       adsurl = {https://ui.adsabs.harvard.edu/abs/2013NatGe...6..661G}
}

@ARTICLE{Rogers2021,
       author = {{Rogers}, James G. and {Gupta}, Akash and {Owen}, James E. and {Schlichting}, Hilke E.},
        title = "{Photoevaporation versus core-powered mass-loss: model comparison with the 3D radius gap}",
      journal = {\mnras},
         year = 2021,
        month = dec,
       volume = {508},
       number = {4},
        pages = {5886-5902},
          doi = {10.1093/mnras/stab2897},
archivePrefix = {arXiv},
       eprint = {2105.03443},
 primaryClass = {astro-ph.EP},
       adsurl = {https://ui.adsabs.harvard.edu/abs/2021MNRAS.508.5886R}
}

@ARTICLE{Ginzburg2018,
       author = {{Ginzburg}, Sivan and {Schlichting}, Hilke E. and {Sari}, Re'em},
        title = "{Core-powered mass-loss and the radius distribution of small exoplanets}",
      journal = {\mnras},
         year = 2018,
        month = may,
       volume = {476},
       number = {1},
        pages = {759-765},
          doi = {10.1093/mnras/sty290},
archivePrefix = {arXiv},
       eprint = {1708.01621},
 primaryClass = {astro-ph.EP},
       adsurl = {https://ui.adsabs.harvard.edu/abs/2018MNRAS.476..759G}
}

@ARTICLE{LehmerCatling2017,
       author = {{Lehmer}, Owen R. and {Catling}, David C.},
        title = "{Rocky Worlds Limited to {\ensuremath{\sim}}1.8 Earth Radii by Atmospheric Escape during a Star{\textquoteright}s Extreme UV Saturation}",
      journal = {\apj},
         year = 2017,
        month = aug,
       volume = {845},
       number = {2},
          eid = {130},
        pages = {130},
          doi = {10.3847/1538-4357/aa8137},
archivePrefix = {arXiv},
       eprint = {1706.02050},
 primaryClass = {astro-ph.EP},
       adsurl = {https://ui.adsabs.harvard.edu/abs/2017ApJ...845..130L}
}

@ARTICLE{OwenWu2013,
       author = {{Owen}, James E. and {Wu}, Yanqin},
        title = "{Kepler Planets: A Tale of Evaporation}",
      journal = {\apj},
         year = 2013,
        month = oct,
       volume = {775},
       number = {2},
          eid = {105},
        pages = {105},
          doi = {10.1088/0004-637X/775/2/105},
archivePrefix = {arXiv},
       eprint = {1303.3899},
 primaryClass = {astro-ph.EP},
       adsurl = {https://ui.adsabs.harvard.edu/abs/2013ApJ...775..105O}
}

@ARTICLE{Hunten1987,
   author = {{Hunten}, D.~M. and {Pepin}, R.~O. and {Walker}, J.~C.~G.},
    title = "{Mass fractionation in hydrodynamic escape}",
  journal = {Icarus},
     year = 1987,
    month = mar,
   volume = 69,
    pages = {532-549},
      doi = {10.1016/0019-1035(87)90022-4},
   adsurl = {http://adsabs.harvard.edu/abs/1987Icar...69..532H}
}

@ARTICLE{Zahnle1990,
   author = {{Zahnle}, K. and {Pollack}, J.~B. and {Kasting}, J.~F.},
    title = "{Mass fractionation of noble gases in diffusion-limited hydrodynamic hydrogen escape}",
  journal = {Icarus},
     year = 1990,
    month = apr,
   volume = 84,
    pages = {502-527},
   adsurl = {http://adsabs.harvard.edu/abs/1990Icar...84..502Z}
}

@ARTICLE{Lincowski2018,
       author = {{Lincowski}, Andrew P. and {Meadows}, Victoria S. and {Crisp}, David and {Robinson}, Tyler D. and {Luger}, Rodrigo and {Lustig-Yaeger}, Jacob and {Arney}, Giada N.},
        title = "{Evolved Climates and Observational Discriminants for the TRAPPIST-1 Planetary System}",
      journal = {\apj},
         year = 2018,
        month = nov,
       volume = {867},
       number = {1},
          eid = {76},
        pages = {76},
          doi = {10.3847/1538-4357/aae36a},
archivePrefix = {arXiv},
       eprint = {1809.07498},
 primaryClass = {astro-ph.EP},
       adsurl = {https://ui.adsabs.harvard.edu/abs/2018ApJ...867...76L}
}

@ARTICLE{Rodriguez2012,
       author = {{Rodr{\'\i}guez}, A. and {Callegari}, N. and {Michtchenko}, T.~A. and {Hussmann}, H.},
        title = "{Spin-orbit coupling for tidally evolving super-Earths}",
      journal = {\mnras},
         year = 2012,
        month = dec,
       volume = {427},
       number = {3},
        pages = {2239-2250},
          doi = {10.1111/j.1365-2966.2012.22084.x},
archivePrefix = {arXiv},
       eprint = {1209.1580},
 primaryClass = {astro-ph.EP},
       adsurl = {https://ui.adsabs.harvard.edu/abs/2012MNRAS.427.2239R}
}

@article{Kopparapu_2013,
doi = {10.1088/0004-637X/765/2/131},
url = {https://dx.doi.org/10.1088/0004-637X/765/2/131},
year = {2013},
month = {feb},
publisher = {The American Astronomical Society},
volume = {765},
number = {2},
pages = {131},
author = {Kopparapu, Ravi Kumar and Ramirez, Ramses and Kasting, James F. and Eymet, Vincent and Robinson, Tyler D. and Mahadevan, Suvrath and Terrien, Ryan C. and Domagal-Goldman, Shawn and Meadows, Victoria and Deshpande, Rohit},
title = {HABITABLE ZONES AROUND MAIN-SEQUENCE STARS: NEW ESTIMATES},
journal = {The Astrophysical Journal}
}

@article{ Magaudda_2020,
	author = {Magaudda, E. and Stelzer, B. and Covey, KR. and Raetz, St. and Matt, SP. and Scholz, A.},
	title = {Relation of X-ray activity and rotation in M dwarfs and predicted time-evolution of the X-ray luminosity⋆},
	DOI= "10.1051/0004-6361/201937408",
	url= "https://doi.org/10.1051/0004-6361/201937408",
	journal = {AI\&A},
	year = 2020,
	volume = 638,
	pages = "A20",
}

@article{ Johnstone_2021,
	author = {Johnstone, C. P. and Bartel, M. and Güdel, M.},
	title = {The active lives of stars: A complete description of the rotation and XUV evolution of F, G, K, and M dwarfs⋆},
	DOI= "10.1051/0004-6361/202038407",
	url= "https://doi.org/10.1051/0004-6361/202038407",
	journal = {AI\&A},
	year = 2021,
	volume = 649,
	pages = "A96",
}

@article{Barnes_2015,
   title={A method to identify the boundary between rocky and gaseous exoplanets from tidal theory and transit durations},
   volume={14},
   ISSN={1475-3006},
   url={http://dx.doi.org/10.1017/S1473550413000499},
   DOI={10.1017/s1473550413000499},
   number={2},
   journal={International Journal of Astrobiology},
   publisher={Cambridge University Press (CUP)},
   author={Barnes, Rory},
   year={2015},
   month={mar}, pages={321–333} 
}

@ARTICLE{Dobrovolskis2009,
       author = {{Dobrovolskis}, Anthony R.},
        title = "{Insolation patterns on synchronous exoplanets with obliquity}",
      journal = {\icarus},
         year = 2009,
        month = nov,
       volume = {204},
       number = {1},
        pages = {1-10},
          doi = {10.1016/j.icarus.2009.06.007},
       adsurl = {https://ui.adsabs.harvard.edu/abs/2009Icar..204....1D}
}

@ARTICLE{Correia2015,
       author = {{Correia}, Alexandre C.~M.},
        title = "{Stellar and planetary Cassini states}",
      journal = {\aap},
         year = 2015,
        month = oct,
       volume = {582},
          eid = {A69},
        pages = {A69},
          doi = {10.1051/0004-6361/201525939},
archivePrefix = {arXiv},
       eprint = {1612.07564},
 primaryClass = {astro-ph.EP},
       adsurl = {https://ui.adsabs.harvard.edu/abs/2015A&A...582A..69C}
}

@ARTICLE{Guerrero2024,
       author = {{Guerrero}, Natalia M. and {Ballard}, Sarah A. and {Su}, Yubo},
        title = "{Plausibility of Capture into High-obliquity States for Exoplanets in the M Dwarf Habitable Zone}",
      journal = {\apj},
         year = 2024,
        month = nov,
       volume = {975},
       number = {2},
          eid = {256},
        pages = {256},
          doi = {10.3847/1538-4357/ad7e28},
archivePrefix = {arXiv},
       eprint = {2310.02332},
 primaryClass = {astro-ph.EP},
       adsurl = {https://ui.adsabs.harvard.edu/abs/2024ApJ...975..256G}
}

@ARTICLE{WilliamsPollard2002,
       author = {{Williams}, Darren M. and {Pollard}, David},
        title = "{Earth-like worlds on eccentric orbits: excursions beyond the habitable zone}",
      journal = {International Journal of Astrobiology},
         year = 2002,
        month = jan,
       volume = {1},
       number = {1},
        pages = {61-69},
          doi = {10.1017/S1473550402001064},
       adsurl = {https://ui.adsabs.harvard.edu/abs/2002IJAsB...1...61W}
}

@ARTICLE{Hut1981,
   author = {{Hut}, P.},
    title = "{Tidal evolution in close binary systems}",
  journal = {Astro.~\& Astrophys.},
     year = 1981,
    month = jun,
   volume = 99,
    pages = {126-140},
   adsurl = {http://adsabs.harvard.edu/abs/1981A%26A....99..126H}
}

@ARTICLE{Heller2011,
   author = {{Heller}, R. and {Leconte}, J. and {Barnes}, R.},
    title = "{Tidal obliquity evolution of potentially habitable planets}",
  journal = {Astro.~\& Astrophys.},
archivePrefix = "arXiv",
   eprint = {1101.2156},
 primaryClass = "astro-ph.EP",
     year = 2011,
    month = apr,
   volume = 528,
    pages = {A27+},
   adsurl = {http://adsabs.harvard.edu/abs/2011A%26A...528A..27H}
}

@ARTICLE{Barnes2007,
       author = {{Barnes}, Jason W.},
        title = "{Effects of Orbital Eccentricity on Extrasolar Planet Transit Detectability and Light Curves}",
      journal = {\pasp},
         year = 2007,
        month = sep,
       volume = {119},
       number = {859},
        pages = {986-993},
          doi = {10.1086/522039},
archivePrefix = {arXiv},
       eprint = {0708.0243},
 primaryClass = {astro-ph},
       adsurl = {https://ui.adsabs.harvard.edu/abs/2007PASP..119..986B}
}

@ARTICLE{Ford2000,
       author = {{Ford}, Eric B. and {Kozinsky}, Boris and {Rasio}, Frederic A.},
        title = "{Secular Evolution of Hierarchical Triple Star Systems}",
      journal = {\apj},
         year = 2000,
        month = may,
       volume = {535},
       number = {1},
        pages = {385-401},
          doi = {10.1086/308815},
       adsurl = {https://ui.adsabs.harvard.edu/abs/2000ApJ...535..385F}
}

@ARTICLE{Agol2005,
       author = {{Agol}, Eric and {Steffen}, Jason and {Sari}, Re'em and {Clarkson}, Will},
        title = "{On detecting terrestrial planets with timing of giant planet transits}",
      journal = {\mnras},
         year = 2005,
        month = may,
       volume = {359},
       number = {2},
        pages = {567-579},
          doi = {10.1111/j.1365-2966.2005.08922.x},
       adsurl = {https://ui.adsabs.harvard.edu/abs/2005MNRAS.359..567A}
}

@ARTICLE{HolmanMurray2005,
       author = {{Holman}, Matthew J. and {Murray}, Norman W.},
        title = "{The Use of Transit Timing to Detect Terrestrial-Mass Extrasolar Planets}",
      journal = {Science},
         year = 2005,
        month = feb,
       volume = {307},
       number = {5713},
        pages = {1288-1291},
          doi = {10.1126/science.1107822},
archivePrefix = {arXiv},
       eprint = {astro-ph/0412028},
 primaryClass = {astro-ph},
       adsurl = {https://ui.adsabs.harvard.edu/abs/2005Sci...307.1288H}
}

@ARTICLE{Dong2018,
       author = {{Dong}, Chuanfei and {Jin}, Meng and {Lingam}, Manasvi and {Airapetian}, Vladimir S. and {Ma}, Yingjuan and {van der Holst}, Bart},
        title = "{Atmospheric escape from the TRAPPIST-1 planets and implications for habitability}",
      journal = {Proceedings of the National Academy of Science},
         year = 2018,
        month = jan,
       volume = {115},
       number = {2},
        pages = {260-265},
          doi = {10.1073/pnas.1708010115},
archivePrefix = {arXiv},
       eprint = {1705.05535},
 primaryClass = {astro-ph.EP},
       adsurl = {https://ui.adsabs.harvard.edu/abs/2018PNAS..115..260D}
}

@ARTICLE{Cohen2015,
       author = {{Cohen}, O. and {Ma}, Y. and {Drake}, J.~J. and {Glocer}, A. and {Garraffo}, C. and {Bell}, J.~M. and {Gombosi}, T.~I.},
        title = "{The Interaction of Venus-like, M-dwarf Planets with the Stellar Wind of Their Host Star}",
      journal = {\apj},
         year = 2015,
        month = jun,
       volume = {806},
       number = {1},
          eid = {41},
        pages = {41},
          doi = {10.1088/0004-637X/806/1/41},
archivePrefix = {arXiv},
       eprint = {1504.06326},
 primaryClass = {astro-ph.EP},
       adsurl = {https://ui.adsabs.harvard.edu/abs/2015ApJ...806...41C}
}

@ARTICLE{Brain2016,
       author = {{Brain}, D.~A. and {Bagenal}, F. and {Ma}, Y. -J. and {Nilsson}, H. and {Stenberg Wieser}, G.},
        title = "{Atmospheric escape from unmagnetized bodies}",
      journal = {Journal of Geophysical Research (Planets)},
         year = 2016,
        month = dec,
       volume = {121},
       number = {12},
        pages = {2364-2385},
          doi = {10.1002/2016JE005162},
       adsurl = {https://ui.adsabs.harvard.edu/abs/2016JGRE..121.2364B}
}

@ARTICLE{Barnes2025lp890,
       author = {{Barnes}, Rory and {do Amaral}, Laura N.~R. and {Birky}, Jessica and {Carone}, Ludmila and {Driscoll}, Peter and {Livesey}, Joseph R. and {Graham}, David and {Becker}, Juliette and {Cui}, Kaiming and {Schlecker}, Martin and {Garcia}, Rodolfo and {Gialluca}, Megan and {Adams}, Arthur and {Ahmed}, MD Redyan and {Bonney}, Paul and {Broussard}, Wynter and {Chawla}, Chetan and {Damasso}, Mario and {Danchi}, William C. and {Deitrick}, Russell and {Ducrot}, Elsa and {Fromont}, Emeline F. and {Gaches}, Brandt A.~L. and {Gupta}, Sakshi and {Hill}, Michelle L. and {Jackman}, James A.~G. and {Janin}, Estelle M. and {Karawacki}, Miko{\l}aj and {Koren}, Matheus Daniel and {La Greca}, Roberto and {Leung}, Michaela and {Miranda-Rosete}, Arturo and {Olohoy}, Michael Kent A. and {Ngo}, Cecelia and {Paul}, Daria and {Sahu}, Chandan Kumar and {Sarkar}, Debajyoti Basu and {Shadab}, Mohammad Afzal and {Schwieterman}, Edward W. and {Sedler}, Melissa and {Texeira}, Katie and {Vazan}, Allona and {Delgado Vega}, Karen N. and {Vijayakumar}, Rohit and {Wojack}, Jonathan T.},
        title = "{History and Habitability of the LP 890-9 Planetary System}",
      journal = {\psj},
         year = 2025,
        month = jan,
       volume = {6},
       number = {1},
          eid = {25},
        pages = {25},
          doi = {10.3847/PSJ/ad94dc},
archivePrefix = {arXiv},
       eprint = {2412.02743},
 primaryClass = {astro-ph.EP},
       adsurl = {https://ui.adsabs.harvard.edu/abs/2025PSJ.....6...25B}
}

@ARTICLE{Airapetian2020,
       author = {{Airapetian}, V.~S. and {Barnes}, R. and {Cohen}, O. and {Collinson}, G.~A. and {Danchi}, W.~C. and {Dong}, C.~F. and {Del Genio}, A.~D. and {France}, K. and {Garcia-Sage}, K. and {Glocer}, A. and {Gopalswamy}, N. and {Grenfell}, J.~L. and {Gronoff}, G. and {G{\"u}del}, M. and {Herbst}, K. and {Henning}, W.~G. and {Jackman}, C.~H. and {Jin}, M. and {Johnstone}, C.~P. and {Kaltenegger}, L. and {Kay}, C.~D. and {Kobayashi}, K. and {Kuang}, W. and {Li}, G. and {Lynch}, B.~J. and {L{\"u}ftinger}, T. and {Luhmann}, J.~G. and {Maehara}, H. and {Mlynczak}, M.~G. and {Notsu}, Y. and {Osten}, R.~A. and {Ramirez}, R.~M. and {Rugheimer}, S. and {Scheucher}, M. and {Schlieder}, J.~E. and {Shibata}, K. and {Sousa-Silva}, C. and {Stamenkovi{\'c}}, V. and {Strangeway}, R.~J. and {Usmanov}, A.~V. and {Vergados}, P. and {Verkhoglyadova}, O.~P. and {Vidotto}, A.~A. and {Voytek}, M. and {Way}, M.~J. and {Zank}, G.~P. and {Yamashiki}, Y.},
        title = "{Impact of space weather on climate and habitability of terrestrial-type exoplanets}",
      journal = {International Journal of Astrobiology},
         year = 2020,
        month = apr,
       volume = {19},
       number = {2},
        pages = {136-194},
          doi = {10.1017/S1473550419000132},
archivePrefix = {arXiv},
       eprint = {1905.05093},
 primaryClass = {astro-ph.EP},
       adsurl = {https://ui.adsabs.harvard.edu/abs/2020IJAsB..19..136A}
}

@article{Barnesetal_2010,
    author = {Barnes, Rory and Jackson, Brian and Greenberg, Richard and Raymond, Sean N.},
    title = {Tidal limits to planetary habitability},
    doi = {10.1088/0004-637X/700/1/L30},
    url={https://iopscience.iop.org/article/10.1088/0004-637X/700/1/L30},
    journal = {The Astrophysical Journal Letters},
    volume = {700},
    number = {1},
    year = {2010}
}

@article{Lee&Owen2025,
	author = {Lee, Eve J.  and  Owen, James E.},
	title = {Carving the Edges of the Rocky Planet Population},
	DOI= "10.3847/2041-8213/adafa3",
	url= "https://doi.org/10.3847/2041-8213/adafa3",
	journal = {The Astrophysical Journal Letters},
	year = {2025},
	volume = {980},
	number = {2},
}

\end{document}